\documentclass[a4paper]{spie}  

\usepackage{amsmath,amsfonts,amssymb}
\usepackage{float}
\usepackage{graphicx}
\usepackage{wrapfig}
\usepackage[colorlinks=true, allcolors=blue]{hyperref}
\usepackage[pagewise]{lineno}
\usepackage{siunitx}        
\DeclareSIUnit\groove{grv}
\DeclareSIUnit\adu{ADU}
\DeclareSIUnit\pixel{pix}
\DeclareSIUnit\electron{\text{\ensuremath{e}}}

\usepackage{geometry}
\newgeometry{               
    left = 1.925cm,
    right = 1.925cm,
    top = 2.54 cm,
    bottom = 3.54cm
}

\title{Sensor characterization for the ULTRASAT space telescope}

\author[1]{Benjamin Bastian-Querner}
\author[1]{Nirmal Kaipachery}
\author[4]{Daniel K\"usters}
\author[1, 2]{Julian Schliwinski}
\author[5]{Shay Alfassi}
\author[1]{Arooj Asif}
\author[1]{Merlin F. Barschke}
\author[3]{Sagi Ben-Ami}
\author[1, 2]{David Berge}
\author[5]{Adi Birman}
\author[1]{Rolf B\"uhler}
\author[1]{Nicola, De Simone}
\author[5]{Amos Fenigstein}
\author[3]{Avishay Gal-Yam}
\author[1]{Gianluca Giavitto}
\author[1]{Juan M. Haces Crespo}
\author[5]{Dmitri Ivanov}
\author[5]{Omer Katz}
\author[1, 2]{Marek Kowalski}
\author[1]{Shrinivasrao R. Kulkarni}
\author[3]{Ofer Lapid}
\author[3]{Tuvia Liran}
\author[3]{Ehud Netzer}
\author[3]{Eran O. Ofek}
\author[1]{Sebastian Philipp}
\author[1]{Heike Prokoph}
\author[6]{Shirly Regev}
\author[3]{Yossi Shvartzvald}
\author[1]{Mikhail Vasilev}
\author[5]{Dmitry Veinger}
\author[1]{Jason J. Watson}
\author[3]{Eli Waxman}
\author[1]{Steven Worm}
\author[1]{Francesco Zappon}
\affil[1]{Deutsches Elektronen-Synchrotron, Platanenallee 6, D-15735 Zeuthen, Germany}
\affil[2]{Institut f\"ur Physik, Humboldt-Universit\"at zu Berlin, Newtonstrasse 15, D-12489 Berlin, Germany}
\affil[3]{Department of particle physics and astrophysics, Weizmann Institute of Science, Herzl St 234, Rehovot, Israel}
\affil[4]{Department of Physics, University of California at Berkeley, 366 LeConte Hall MC 7300, Berkeley, CA 94720-7300, USA}
\affil[5]{Tower Semiconductor, 20 Shaul Amor Avenue, Migdal Haemek, 2310502, Israel}
\affil[6]{Etesian Semiconductor Ltd, P.O.B 3227, Ramat Yishai, 3004205, Israel}

\authorinfo{Send correspondence to Julian Schliwinski, E-mail: julian.schliwinski@desy.de}

\begin{document} 
\maketitle

\begin{abstract}
    The Ultraviolet Transient Astronomical Satellite (ULTRASAT) is a scientific space mission carrying an astronomical telescope. The mission is led by the Weizmann Institute of Science (WIS) in Israel and the Israel Space Agency (ISA), while the camera in the focal plane is designed and built by Deutsches Elektronen Synchrotron (DESY) in Germany. Two key science goals of the mission are the detection of counterparts to gravitational wave sources and supernovae\cite{Sagiv_2014}. The launch to geostationary orbit is planned for 2024. The telescope with a field-of-view of $\approx200\,$deg$^2$, is optimized to work in the near-ultraviolet (NUV) band between $220$ and $280\,$nm. The focal plane array is composed of four $22.4$\nobreakdash-megapixel, backside-illuminated (BSI) CMOS sensors with a total active area of $90\times 90\,$mm$^2$\cite{Asif_2021}. Prior to sensor production, smaller test sensors have been tested to support critical design decisions for the final flight sensor. These test sensors share the design of epitaxial layer and anti-reflective coatings (ARC) with the flight sensors. Here, we present a characterization of these test sensors. Dark current and read noise are characterized as a function of the device temperature. A temperature-independent noise level is attributed to on-die infrared emission and the read-out electronics' self-heating. We utilize a high-precision photometric calibration setup\cite{Kuesters2020} to obtain the test sensors’ quantum efficiency (QE) relative to PTB/NIST-calibrated transfer standards ($220$\nobreakdash-$1100\,$nm), the quantum yield for $\lambda < 300\,$nm, the non-linearity of the system, and the conversion gain. The uncertainties are discussed in the context of the newest results on the setup’s performance parameters.
    From three ARC options, Tstd, T1 and T2, the latter optimizes out-of-band rejection and peaks in the mid of the ULTRASAT operational waveband (max. QE $\approx 80\,\%$ at $245\,\mathrm{nm}$) . We recommend ARC option T2 for the final ULTRASAT UV sensor.
\end{abstract}


\newpage

\section{INTRODUCTION}
\label{sec:introduction}

    ULTRASAT is a space mission instrumented with a scientific telescope which is designed for time domain astronomy.
    
    Under the mission leadership of Weizmann Institute of Science (WIS)\footnote{\href{https://www.weizmann.ac.il/particle/}{Weizmann Institute of Science}, 234 Herzl Street, Rehovot 7610001 Israel} and Isreal Space Agency (ISA)\footnote{\href{https://www.space.gov.il/en/}{Israel Space Agency}, Derech Menachem Begin 52, Tel Aviv, Israel}, the project responsibilities are shared among science institutes and industry partners. The camera in the telescope focal plane is designed and developed by DESY\footnote{\href{https://www.desy.de}{Deutsches Elektronen-Synchrotron DESY}, Platanenallee 6, 15738 Zeuthen}. The launch to geostationary orbit is planned the second half of 2024.
    
    The distinguishing feature of ULTRASAT and its Schmidt telescope is the wide field of view (FoV) of $\approx200$ square degrees. It will perform repeated observations of the sky with $300\,\text{s}$ cadence in the near ultra violet waveband (NUV, $220\,\textnormal{-}280\,\text{nm}$).
    ULTRASAT will be capable of a high detection rate of transient events to enable the detection of EM counterparts of gravitational wave sources, tidal disruption events, in addition to supernovae~\cite{Sagiv_2014}.
    
    The central element of the telescope's camera is the detector assembly equipped with four independent BSI CMOS UV sensor tiles. Each tile provides a photosensitive area of $45\times45\,\text{mm}^2$ consisting of $9.5\times9.5\,\mu\text{m}^2$ pixels. The pixels are realised in a 5T-design, that offers dual gain capability enabling a high-dynamic range operation mode. In total, the four sensor tiles have 89.8 Megapixel. The overall design of the camera has passed the preliminary design review and first models are expected in 2022\cite{Asif_2021}.

    \begin{figure} [ht]
       \begin{center}
           \begin{tabular}{cc}
               \includegraphics[height=6.5cm]{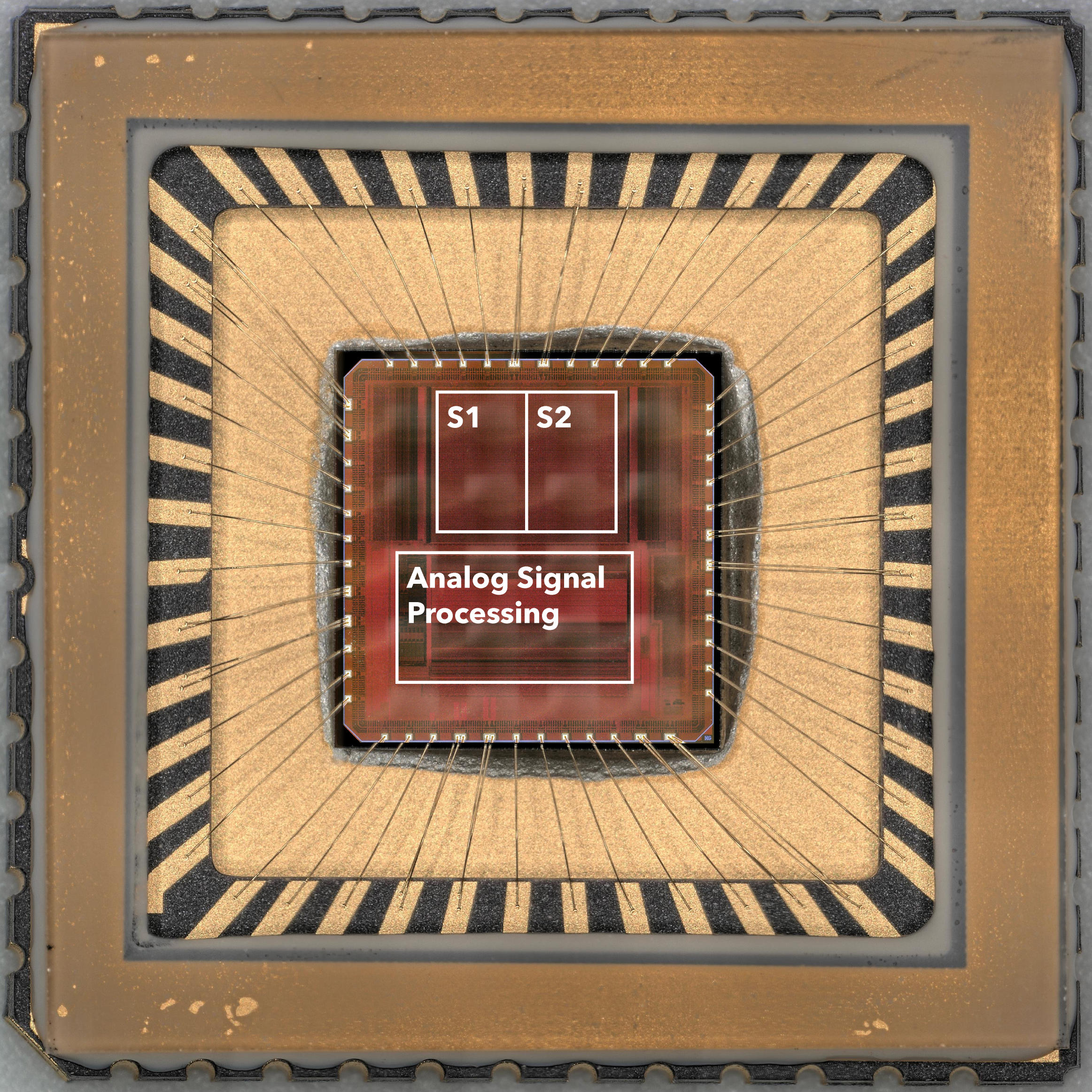} & \includegraphics[height=6.5cm]{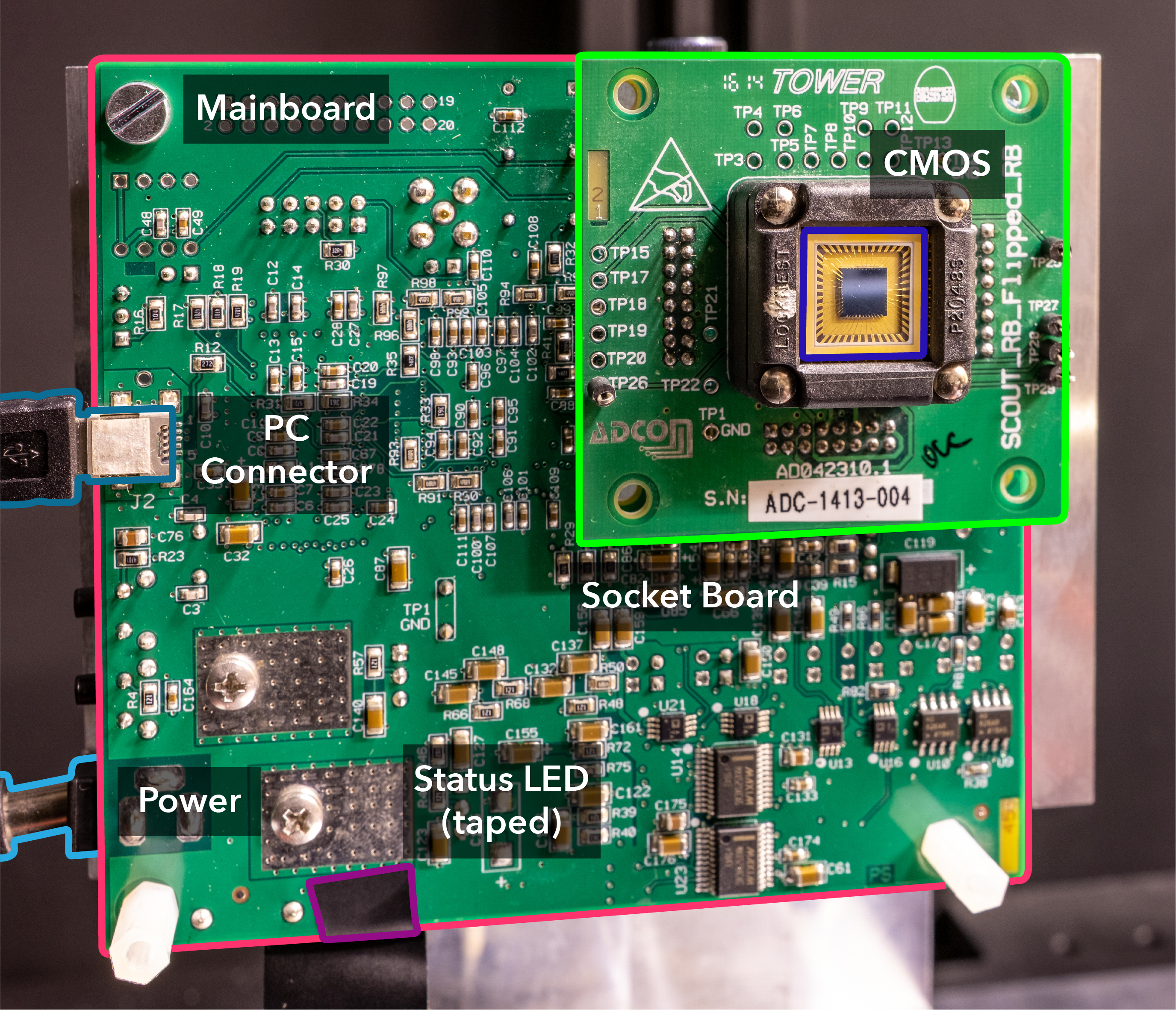} 
           \end{tabular}
        \end{center}
        \caption[example] 
        {\label{fig:sens_pack_devboard} 
        The test sensor package is shown (left), highlighting the two active pixel arrays, labeled S1 and S2, and the analog signal processing region on the central BSI CMOS die. The test sensor features a 4T pixel design with fixed gain and $15\,\mu\text{m}$ pitch. The two pixel arrays differ in one respect, that is the additional N-Well inside S2 operating as a guard ring for each pixel to reduce blooming effects. For ULTRASAT, however, there is no such guard ring planned. Hence, if not stated differently, only pixel array S1 is subject to our measurements reported here.  The development board is shown (right) in operational condition with mounted sensor and placed inside the light-tight laboratory environment. The different components and interfaces of the system are marked in the picture.}
    \end{figure}
    \begin{figure} [ht]
       \begin{center}
           \begin{tabular}{c}
               \includegraphics[height=6.5cm]{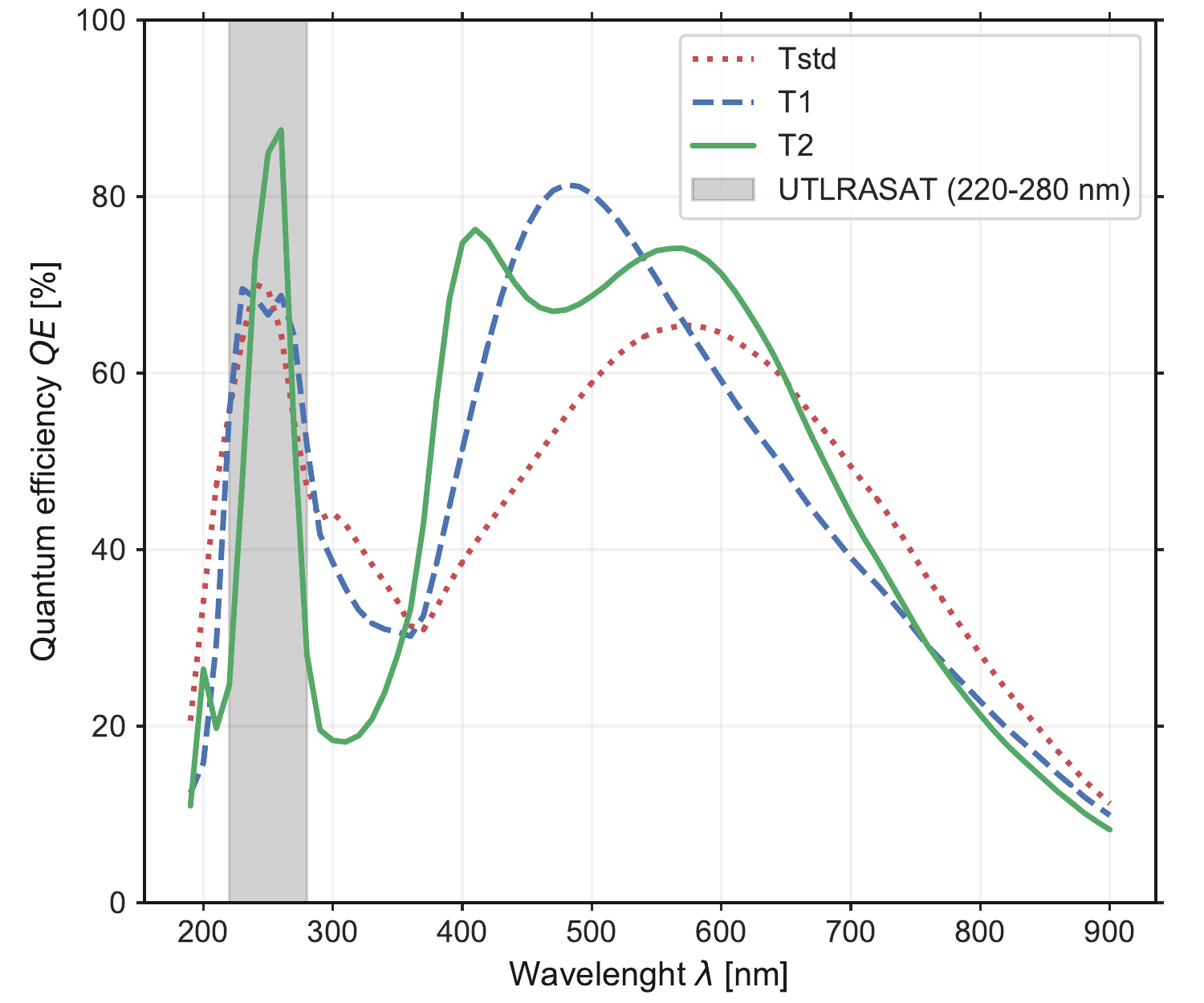}
           \end{tabular}
        \end{center}
        \caption[example] 
        {\label{fig:qe_sim} 
        Numerical simulations comparing the three different ARC options, labeled Tstd, T1 and T2, show the quantum efficiency of the test sensors for light incidence at an angle of $25$ degrees. The ULTRASAT operational waveband is highlighted in gray over the range of $220\,\text{nm} \leq \lambda \leq 280\,\text{nm}$.}
    \end{figure}
    In advance of production of the final ULTRASAT flight sensor, test sensors  sharing a comparable pixel design and epitaxial layer were provided by Tower Semiconductors (Fig.~\ref{fig:sens_pack_devboard}). We report on the characterization study of these test sensors which is carried out in order to inform the decision on the type of anti-reflective coating (ARC, Fig.~\ref{fig:qe_sim}) to be used on the final sensor. Furthermore, this study aims to verify whether the test sensors satisfy the ULTRASAT performance requirement for noise characteristics at the operating temperature of $200\,\text{K}$. The measurements rely on an updated high-precision photometric calibration setup~\cite{Kuesters2020}. The characterization encompasses a comprehensive test of sensor gain, linearity, dark current as well as quantum yield and quantum efficiency. 

\section{PHOTOMETRIC CALIBRATION}
\label{sec:photometric_calibration}
    The photo-metric calibration setup at DESY Zeuthen provides monochromatic light from $200\,$nm to $1200\,$nm. As a light source, we use a Laser-Driven Light Source (LDLS) which supplies a broadband spectrum with enhanced output in the UV. The light is dispersed by two monochromators working in serial with blazed diffraction gratings. The combined bandwidth is $1.7\, $nm or $3.4\, $nm depending on the chosen gratings. The light flux on the test sensor is estimated during its illumination. A photodiode measures one-half of the setup's light. It works as our working standard (WS) which is calibrated against our two flux reference photodiodes or primary standard (PS) calibrated by the Physikalisch Technischen Bundesanstalt (PTB)\footnote{Calibrated from $200\;$nm to $400\;$nm.} and the National Institute of Technology and Standardisation (NIST)\footnote{Calibrated from $300\;$nm to $1100\;$nm.}. As a wavelength standard, we use low-pressure gas lamps\footnote{The wavelength of the emission lines are taken from the Atomic Spectra Database\cite{Nist_spec}}. To transfer the calibration from the spectral lines to the LDLS's spectrum, we use an absorption line filter.  

    \subsection{Laboratory setup}
    \label{sub:photometric_setup}
    
\begin{figure}[]
            \begin{center}
                \includegraphics[width=0.85\textwidth]{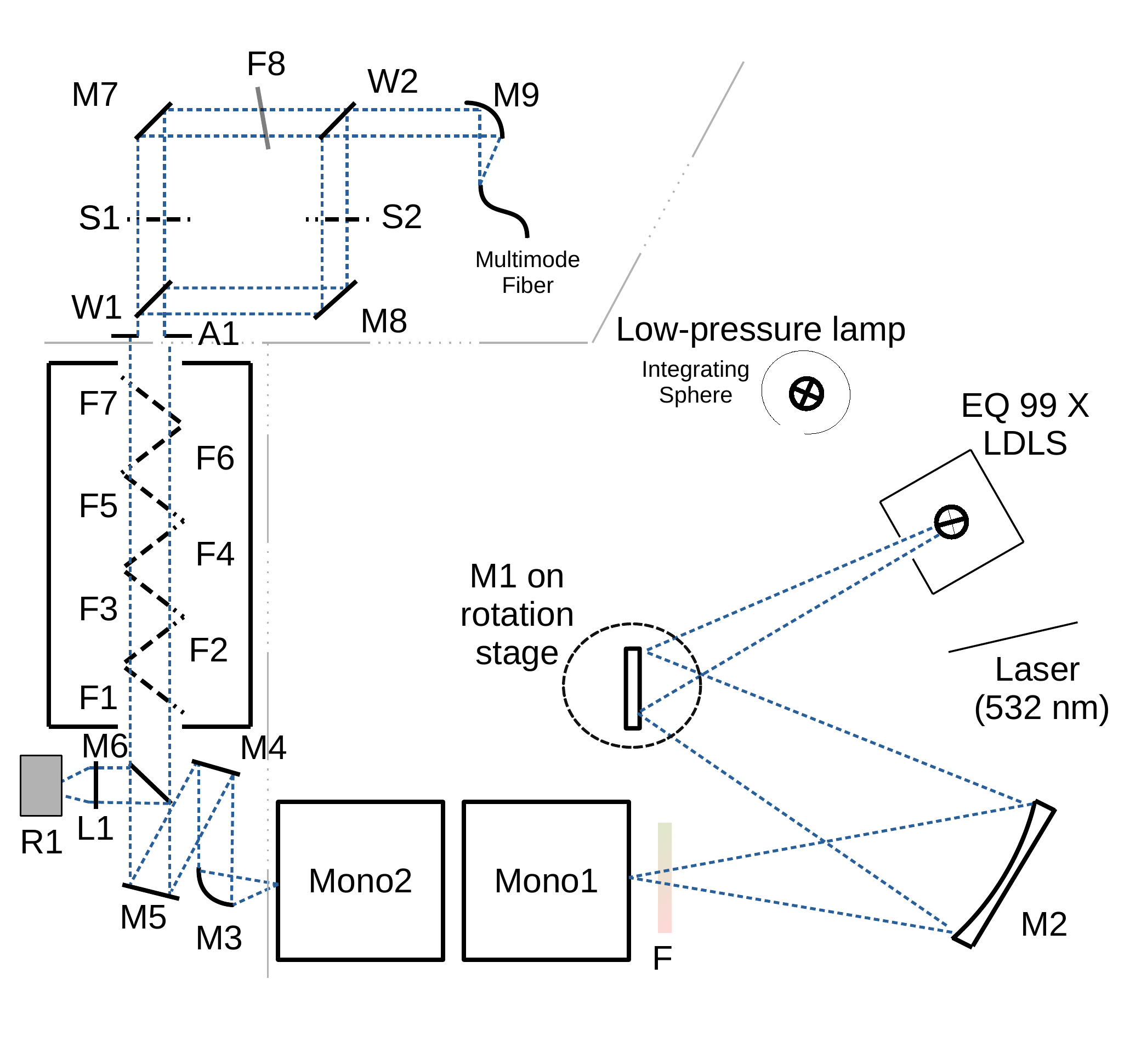}
                \caption[example] 
                {\label{fig:setup_scheme} 
                Schematic overview of the light source section of the calibration setup.
                  The components are:\newline
                  - Lamps: LDLS, 532\,nm Laser, low pressure gas lamp Hg\,/\,Ne\,/\,Ar\,/\,Na\,/\,Cd\newline
                  - M1: Flat mirror, rot-able (mounted on HDR50/M from Thorlabs) to select light sources             \newline
                  - M2: Spherical mirror, $f=0.4$\,m, $d=200$\,mm \newline
                  - F: Filter wheel with longpass filters.\newline
                  - Mono1, Mono2: Cornerstone 260 monochromators\newline
                  - M3: Off-axis parabolic mirror from Thorlabs, $d=1$", ${\mathrm{RFL}}=2$"\newline
                  - M4: Concave cylindrical mirror, $d=1$", $f=2$"\newline
                  - M5: Flat folding mirror\newline
                  - M6: UV fused silica window as a beam splitter to retrieve reference light beam\newline
                  - A1: Aperture to define beam diameter\newline
                  - F1-F7: Reflective neutral density filters with ${\mathrm{ OD}}=0,1,2$\newline
                  - W1/2: UV fused silica windows $d=2$", used as beam splitter / combiner \newline
                  - S1/2: shutters operated \newline
                  - M7,M8: remote controlled shutters\newline
                  - F8: Continuously variable reflective neutral density filter ${\mathrm{OD:}}\,0.04-2.0$ mounted to a HDR50 rotation stage\newline
                  - M9: Off axis parabolic mirror, same as M5, with a multimode fiber in its focus to the gantry robot, type FG400AEA.}
         	\end{center}
        \end{figure}

        A flat mirror (M1, see Fig.~\ref{fig:setup_scheme}), which is mounted on a Thorlabs HDR50 rotation stage, reflects the light from different sources to illuminate the spherical mirror (M2\footnote{spherical mirror of 40\,cm and a diameter of 20\,cm.}). In the case of the LDLS (EQ99-X from Energetiq), two Thorlabs 2-inch off-axis parabolic mirrors transform an F/1 beam to an F/4 beam fitting the monochromator. A laser diode (Thorlabs CPS532-C2 laser diode at 532\,nm with 0.9\,mW) is available for alignments. The low-pressure gas lamps are encapsulated in a PTFE cylinder, acting as an integrating sphere with an output slit of $\approx 5\,$mm width. M2 focuses the light onto the entrance slit of the first monochromator. 
        Before it enters the monochromator it passes a filter wheel with longpass filters (F)\footnote[1]{No filter: $\lambda$\,$<$\,320\,nm; WG280: 320\,nm\,$<$\,$\lambda$\,$<$\,500\,nm; GG495: 500\,nm\,$<$\,$\lambda$\,$<$\,750\,nm; RG695: 750\,nm\,$<$\,$\lambda$\,$<$\,870\,nm; RG830: 870\,nm\,$<$\,$\lambda$}. These longpass filters remove the light from the input broadband source that otherwise would be transmitted as higher-order light due to the grating's interference. It is necessary to position these filters in front of the monochromator to remove the unavoidable fluorescent emission they produce \cite{Reichel_2015}. 
         
        We use two Oriel Cornerstone 260 monochromators from Newport, whereby the entrance slit of the second monochromator replaces the exit slit of the first monochromator. In this configuration, the first monochromator preselects the wavelength range, and the second suppresses the out-of-band light from the first. Each grating turret is connected to Heidenhain ERN 480 5000 high-resolution encoders, allowing sub-arcsecond angle measurements to improve the system's wavelength calibration. For a detailed description of the double monochromator and the calculation of the system's wavelength, see \cite{Kuesters2020}. The beam leaving the double monochromator is collimated using a Thorlabs off-axis parabolic mirror (M3). To account for astigmatism from the double monochromator, we use a cylindrical mirror (M4) and redirect the beam into the attenuator setup with a flat mirror (M5). 
         
        The beam passes a UV-fused silica window (M6) used as a beam splitter. The reflected beam is focused by a UV-fused silica lens (L1) on a reference diode to track the variations in intensity. Using a system of seven linear motors F1-7, we can insert five reflective neutral density\footnote[2]{Three reflective neutral density filters with nominal optical density (OD) of one, and two filters with the nominal optical density of two, from Edmund optics.} filters and two clear UV-fused silica windows into the light beam. The filters are inserted at an angle of $\pm45$\,$^{\circ}$ to reject either $90\%$ or $99\%$ from the beam. The remaining light is split into two beams (W1\footnote[3]{W1, W2: UV fused silica windows $d=2$", used as beam splitter/combiner}), which are recombined with W2 after each one passing a shutter (S1 \& S2). The recombined beam is then focused by an off-axis paraboloid (M9) onto a multimode fiber\footnote[4]{Mounted on a motorized 2d stage to enhance coupling efficiency.}. We can perform measurements with beam A (shutter S1), beam B (shutter S2), or both beams using the two shutters. If the test sensor is linear, adding the measured value with beam A and the measured value with beam B should equal the measured amount with both beams. A round variable neutral density filter (F8) allows us to decrease the light intensity for beam A by two additional optical densities continuously. 
         
\begin{figure} [ht]
            \begin{center}
               \begin{tabular}{cc}
                   \includegraphics[height=6.5cm]{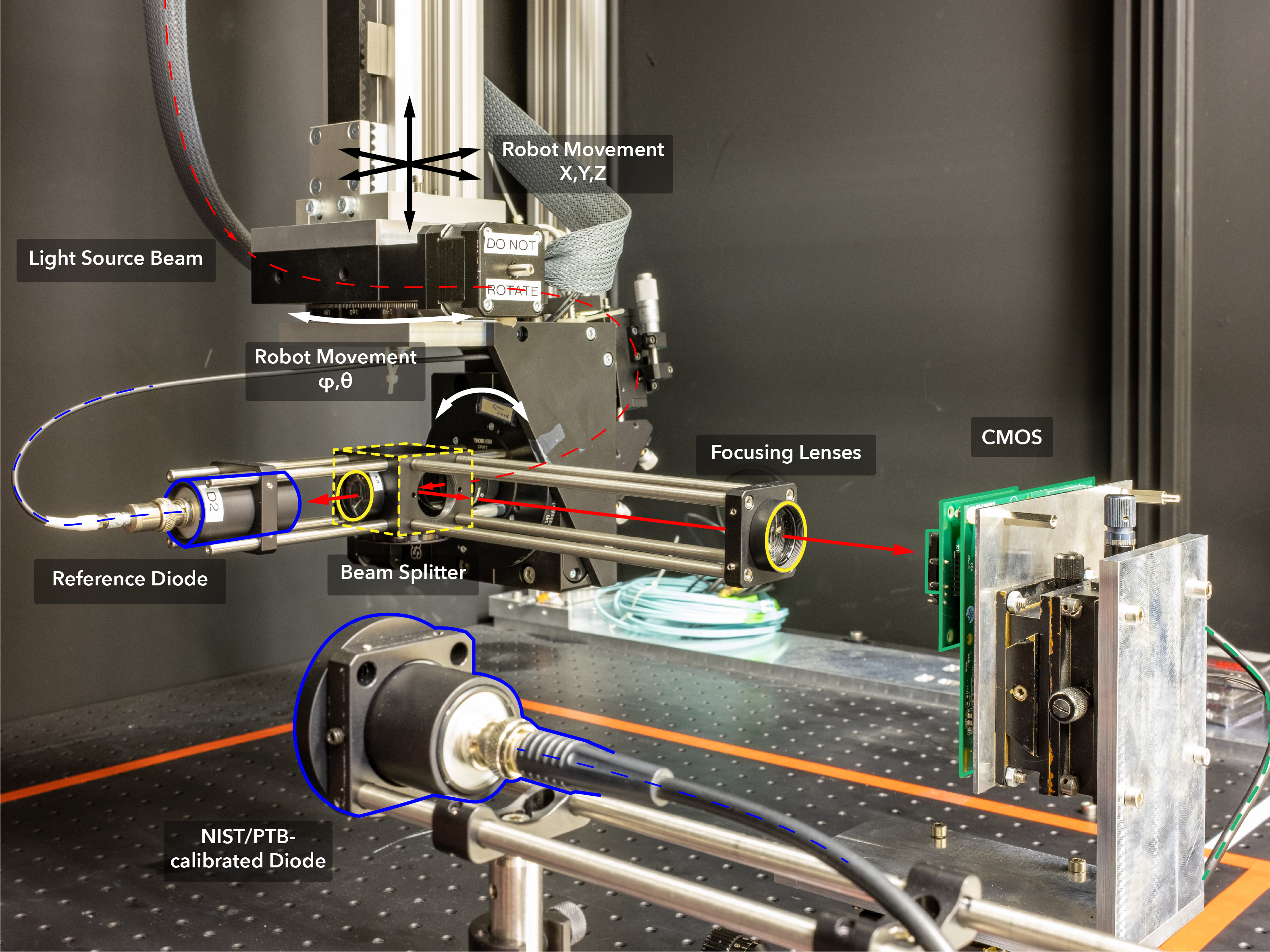} & \includegraphics[height=6.5cm]{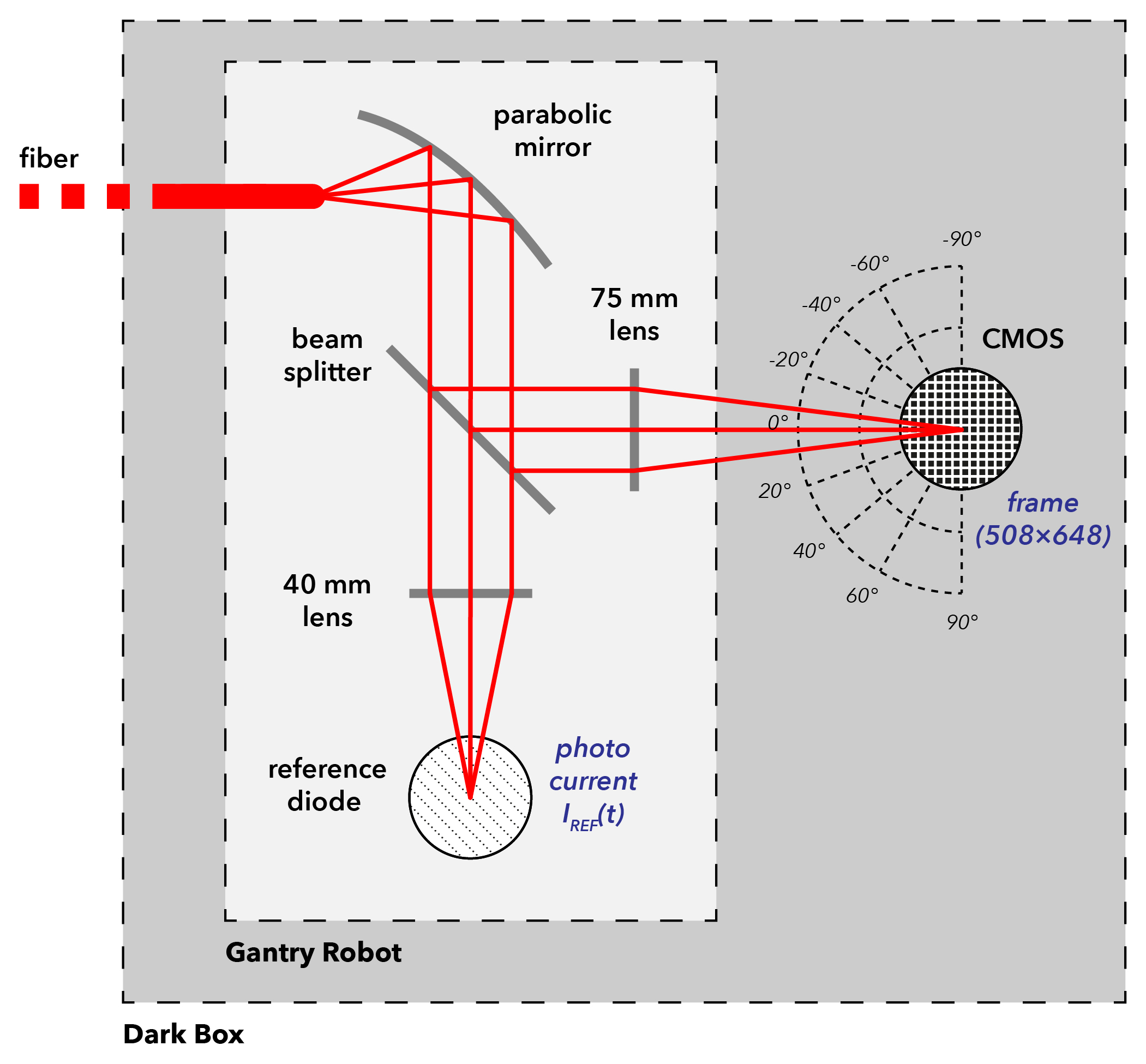}
               \end{tabular}
            \end{center}
            \caption[example] 
            {\label{fig:qe_setup} 
            The picture (left) shows the inside of the light-tight optical laboratory with the most important components highlighted. Through a multimode fiber, the output from the light source enters the setup, where it is mounted on a remote-controlled five-axis gantry robot. The stage carries a collimating mirror and a beam splitter from which light is focused onto a photodiode, operating as a working standard, and the test sensor. The schematic (right) of the setup illustrates the beam path. The gantry robot can move relative to the test sensor. At focus distance ($75\,$mm), the spot's FWHM is about $(530\pm30)\,\mathrm{\mu m}$ which covers approx. $1000$\,pixels. The test sensor can be rotated around its vertical axis for a variation of the incident angle.}
        \end{figure}

        The test sensor is measured in a light-tight enclosure. The light from the source enters the enclosure and is transported via a multimode fiber of type FG400AEA (18\,m length) to the final optics held by a five-axis gantry robot (see Fig.~\ref{fig:qe_setup}, left). The robot can change its spatial position in all three Cartesian axes $[0,1070]\,\mathrm{mm}\times[0,680]\,\mathrm{mm}\times[0,580]\,\mathrm{mm}$ and rotate the optics $360\,^\circ$ in azimuthal and polar axes within $[70, 180]\,^\circ$. A reflective neutral density filter is used as beam splitter (W3) to probe $\approx 50\,\%$ of the fiber output flux using a Hamamatsu S1337-1010BQ diode as WS (see Fig.~\ref{fig:qe_setup}, right). The reflected part is used to be focused onto the test sensor. The WS's spectral response is calibrated together with the splitting factor of W3 by comparing it with the response of our PS placed as a test sensor. The light spot on the test sensor has an FWHM of $(530\pm30)\,\mathrm{\mu m}$. The light spot is always fully contained within the collecting area $A_c$ of the diode or the pixel array of the test sensor. A more detailed description about the setup and its characteristics can be found in K\"usters et al.\cite{Kuesters2020}.
        
    \subsection{System performance}
    \label{sec:performance_results}
        The resulting light flux on the test sensor is a product of the spectral power of the LDLS, the throughput of the longpass filter, the spectral efficiency of the gratings, the throughput of the multimode fiber, the combined efficiency of the mirror optics, as well as the additional chosen optical density with the reflective neutral density filters. 
        We use pairs of blazed diffracting gratings. For wavelengths in the visual (VIS) and infrared (IR) bands ($\lambda > 365\,$nm) the number of grooves per mm of the gratings is \SI{1200}{\groove\per\milli\meter} whereas in the ultraviolet (UV) ($\lambda \le 365\,$nm) it is smaller at \SI{600}{\groove\per\milli\meter} to increase the light intensity in this regime. Resulting from that, the overall bandwidths are  $\approx 1.7\,$nm in the VIS/IR  and $\approx 3.4\,$nm in the UV. Moreover, the generated photocurrent at the working standard varies depending on the spectral range. In the UV it ranges from 1\,pA at 220\,nm to 200\,pA at 365\,nm and in the VIS/IR bands it varies between 1\,pA and 1\,nA.
        
        \paragraph{Uncertainty on flux scale}
        We use a Dual-Channel Picoammeter 6482 from Keithley to transfer the flux calibration from the PS to our WS and estimate the light flux on the test sensor. Our primary standards are Hamamatsu S1337-1010BQ (ultraviolet, calibrated by PTB) and S2281 (optical and infrared, calibtrated by NIST) photodiodes. The calibrations have a maximal uncertainty of $0.3\,\%$ from PTB and $0.15\,\%$ from NIST, respectively. The statistical errors dominate our uncertainties in the photocurrent.  The uncertainty for the absolute value is estimated from the manufacturers calibration to be $0.2\,\%$ whereby the statistical error lies between $0.1\,\%$ for $1\,$nA and $2\,\%$ for $1\,$pA. As the generated photocurrent in ultraviolet lies between $1\,$pA and $100\,$pA the statistical uncertainty is the dominating error in the ULTRASAT wavelength regime.
        
        \paragraph{Uncertainty on the wavelength scale}
        The used monochromators have a periodic error in the drive of the grating turret and we observed stepping losses of their stepper motors. Using a high-resolution encoder mounted to the grating turret, we measure the grating angles, which allows us to predict the wavelength of the light passing the monochromator system\cite{Kuesters2020}. To verify the predicted wavelength or model wavelength, we compare the measured wavelength of emission lines in the spectra of low-pressure gas lamps to the NIST database\cite{Nist_spec}. To find the "right" wavelength in the NIST database, we use a loaned Fourier transform spectrometer from Thorlabs (OSA201C, resolution of $\le$\,20\,pm for wavelength $\le$\,1\,$\mu$m) to assign the emission lines of our lamps to the NIST database. We now obtain synthetic spectra for the lamps and transfer the NIST wavelengths to our model by fitting local deviations between synthetic and predicted wavelength. These deviations follow a linear relation in the model wavelength.
   
        For each low-pressure lamp, we find a different linear calibration, with a spread of $1\,$\AA, whereby the residuals for each lamp fit are in the order of \textcolor{black}{$\approx0.2\,$\AA}. We attribute the difference between the lamps to inhomogeneity in the illumination of the first monochromator's entrance slit, caused by direct light from the lamp\footnote[6]{The PTFE cylinder around the lamp is of insufficient size.}. This linear trend can be calibrated out. However the calibration with the emission lines is only applicable to light from the PTFE cylinder and can not be transferred to the spectrum of the LDLS directly. Instead, we measure the spectral transmission of a Holmium Didymium absorption line filter (HoDi Filter) from Hellma, type UV45, to obtain the systematic uncertainty on the model wavelength with the LDLS. The filter provides absorption lines between $241\,$nm and 864nm with optical densities between 0.07 and 0.74. It is shipped with a calibration provided by the manufacturer\footnote[7]{Its laboratory is approved by the Deutsche Akkreditierungsstelle (DAkkS).} with an accuracy to 0.2\,nm. 
    
        To determine the system's stability, we performed repeated scans of three emission lines, the first line, a blend line from two CdI lines at 325.34622\,nm, and 326.19951\,nm, the second, a HgI line at 546.22675\,nm and third, a HgI line at 1014.253\,nm. We scanned for changes within $\approx$\,300 hours or $\approx$\,70 repetitions with varying laboratory temperature. We find a correlation between the temperature and the fitted line centers within a temperature range from $17^{\circ}\,$C to $24^{\circ}$ and a linear temperature coefficient of $9.7\,$pm / $^{\circ}$C for all wavelengths.
        \par
        We have shown that our double monochromator system has a high reproducibility and can be calibrated to high accuracy in the case of a homogeneous illumination of the monochromator entrance slit.\cite{Kuesters2020} Applying the linear calibration with the HELLMA filter back to the QE measurements would require prominent signatures in the lamp spectrum. The spectrum of the LDLS shows weak emission lines in the VIS and the strong lines in the IR. A calibration on these would result in large extrapolation errors. We quantified the influence of the inhomogeneous illumination of the entrance slit by comparing filter calibrations with different  couplings. The systematic uncertainty in the QE and QY measurements for the test sensors is estimated as $\le$\,2\,nm and dominating those from the varying laboratory temperatures and measurement statistics.
        
        To ensure a homogeneous illumination of the monochromator, we will implement a hollow PTFE cylinder of increased size to encapsulate the low-pressure gas lamp and geometrically avoid direct light reaching the monochromator. The new cylinder will also contain a borehole to illuminate the cylinder with a Xe high-pressure arc lamp. In this way, we can transfer the linear calibration of our model wavelength to the absorption lines of the HoDi Filter with the precision of the emission line fits and apply this to the measurements using the LDLS. This will allow us to decrease the wavelength uncertainty for the ULTRASAT flight sensor's measurements and other projects to $\le$\,0.4\,\AA.

\section{SENSOR CHARACTERIZATION}
\label{sec:sensor_characterization}
    
    To asses the question if the test sensors design fulfill the project's requirements and to achieve a recommendation of the ARC options we perform several tests on the test sensors provided by Tower Semiconductors. The test sensors are characterized in their gain, the read noise, the dark current as well as the spectral quantum efficiency. Further, we quantify the detector's quantum yield and investigate it in its non-linearity.

    \subsection{Gain, non-linearity and leakage current}
    \label{sec:gain_non_linearity_and_leakage_current}
        The gain of CMOS sensors refers to the conversion factor used to translate the voltage generated by the collected charge in each CMOS pixel into units of the analog-to-digital converter (ADU). The gain $K$ is expressed in ADUs per electron (\si{\adu\per\electron}) and is the constant of proportionality between the mean number of collected electrons $\mu_e$ and the mean digitized signal $\mu_{digi}$ of the sensor. Hence, the linear relationship is written as $\mu_{\text{digi}} = K \cdot \mu_e$.

        The knowledge of gain plays a crucial role in measuring the absolute quantum efficiency where the number of photoelectrons detected by the test sensor is compared to the number of photons estimated from the working standard's photocurrent under investigation.

        The gain is characteristic to components of the sensor's analog signal processing circuits and can be determined from direct measurements on these components. The manufacturer reports a conversion factor of $52\,\mu\text{V}/\text{e}$. This is related to the gain $K \hat{=} N_{\text{bit}} / V_{\text{AD}} \cdot 52\,\mu\text{V}/\text{e} = 0.42\,\text{ADU}/\text{e}$, with a bit depth of $N_{\text{bit}} = 2^{14}$ and a supply voltage range $V_{\text{AD}}$ of the analog-to-digital converter of $2\,\text{V}$.
        
        We use an alternative method to determine the gain which exploits the statistical nature of the signal found in sensor images recorded under exclusion of light. To separate between sensor images with and without incident light, we refer to light and dark images, respectively. Our procedure is comparable to the photon transfer method (PTM)~\cite{Janesick1987}. In this case, gain is determined from the analysis of dark images only. More specifically, an exposure time sequence of dark images is conducted. That is, several dark images are recorded at different durations of exposure time, $t_{\text{exp}}$. The timescale of these exposures reaches from one microsecond up to several seconds.
        
\begin{figure} [ht]
            \begin{center}
               \begin{tabular}{c}
                   \includegraphics[height=6.5cm]{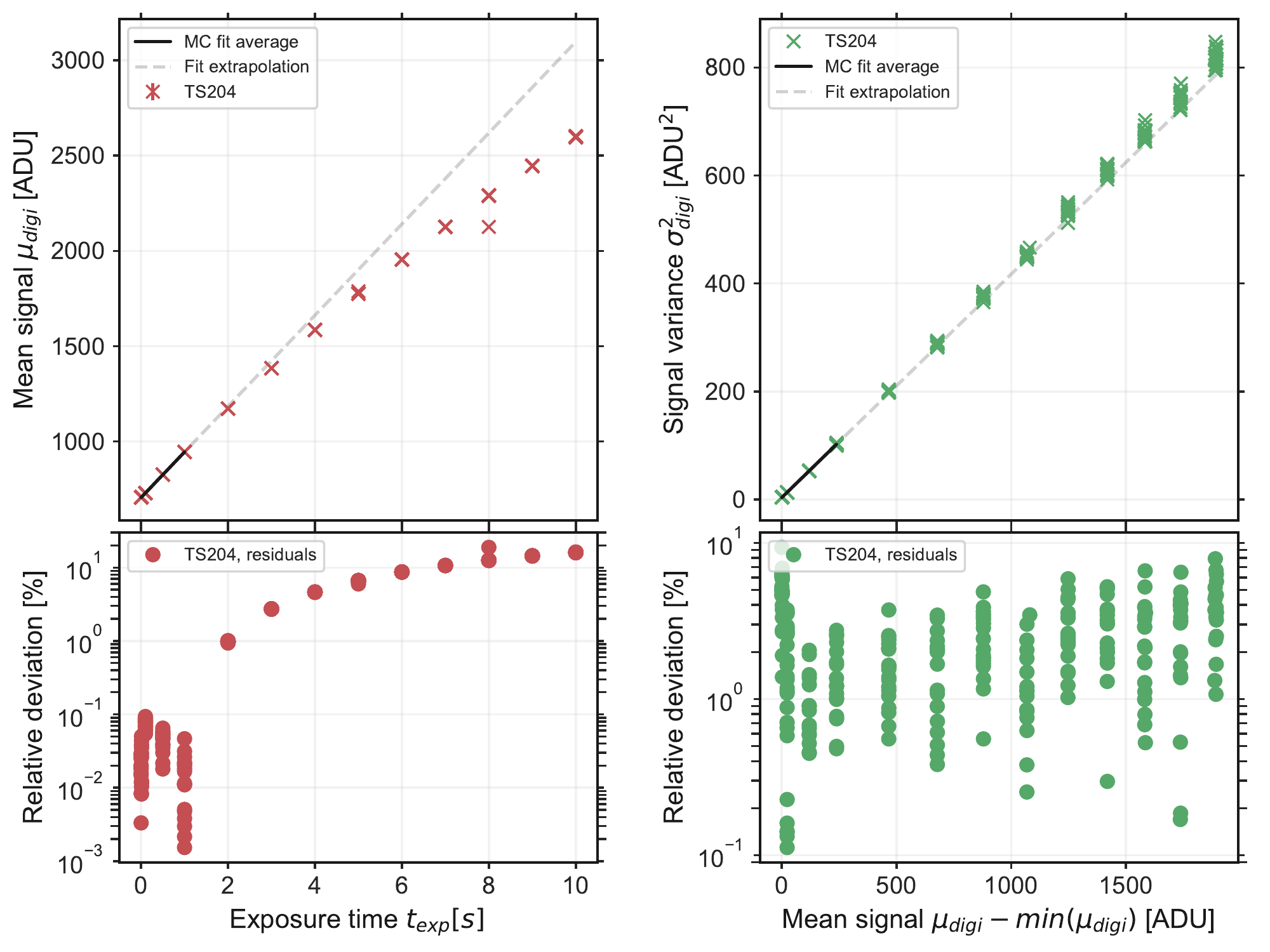}
               \end{tabular}
            \end{center}
            \caption[example] 
            {\label{fig:dc_lin_fit} 
            The exemplary display the gain measurement results of the test sensor with ID \emph{TS204} shows the mean signal against exposure time (left) and signal variance against bias-subtracted mean signal. The residuals are shown as relative deviation from the linear best-fit model in the lower panels. Each marker represents the data corresponding to an individual image pair.}
        \end{figure}

        We first measure pairs of images with same exposure times which are considered  to eliminate fixed pattern noise. Note, that the paring of images is exclusive for any given pair such that correlation between pairs is avoided. From the image pair average, the mean signal per pixel $\mu_{\text{digi}}$ is calculated. Whereas the signal variance per pixel $\sigma_{\text{digi}}^2$ is estimated from the image pair difference. Second, the mean signal per pixel from all available image pairs is plotted against exposure time (Fig.~\ref{fig:dc_lin_fit}, left panel) whereby we expect a linear behavior for ideal sensors. This relationship is characterized by the dark current rate $DCR$ (in ADU per pixel per second) and the offset given by the test sensor's bias level $\mu_{\text{bias}}$. A best-fit estimate is used to determine $\mu_{\text{bias}}$ and $DCR$. Finally, the signal variance per pixel is plotted against the mean signal per pixel (Fig.~\ref{fig:dc_lin_fit}, right panel) for all image pairs. Based on the assumption that the number of collected electrons follows Poisson statistics\cite{Canfield_1998}, the linear relationship is given by the gain $K$ as proportionality constant and y-axis offset by the test sensor's read noise $\sigma_{\text{read}}^2$.
        
        Note, that significant deviations from linear behavior have been observed in both cases for higher values of exposure time and mean signal. Possible effects that could account for this are discussed towards the end of this section. However, the inference of a model with which corrections, i.e. re-linearization of the measured data~\cite{soman2015non}, could be achieved was found to be outside the scope of this study. In order to minimize the impact by any non-linearities, the processing of all data sets is limited below a maximum of $1\,\text{s}$ and $2000\,\text{ADU}$, respectively. To better account for the statistical fluctuations and especially the systematics introduced by remaining non-linearities, we used Monte-Carlo-simulated data based on the initial measurement data to estimate the best-fit parameters. The mean value and standard deviation of the resulting distribution of best-fit parameters are taken as the final measurement result
        
        Throughout this study, we evaluate a total of twelve test sensors. Thus, a central gain value can be calculated from this sample to be used in the analysis of subsequent measurements. Additionally, the analysis procedure yields the test sensor's read noise as a side product and, similarly, a central value can be calculated. These values are presented in Tab.~\ref{tab:dc_results}. For the discussion of the dark current results, the reader is referred to Section~\ref{sec:dark_current}.
        \begin{figure} [ht]
            \begin{center}
               \begin{tabular}{cc} 
               \includegraphics[height=6.7cm]{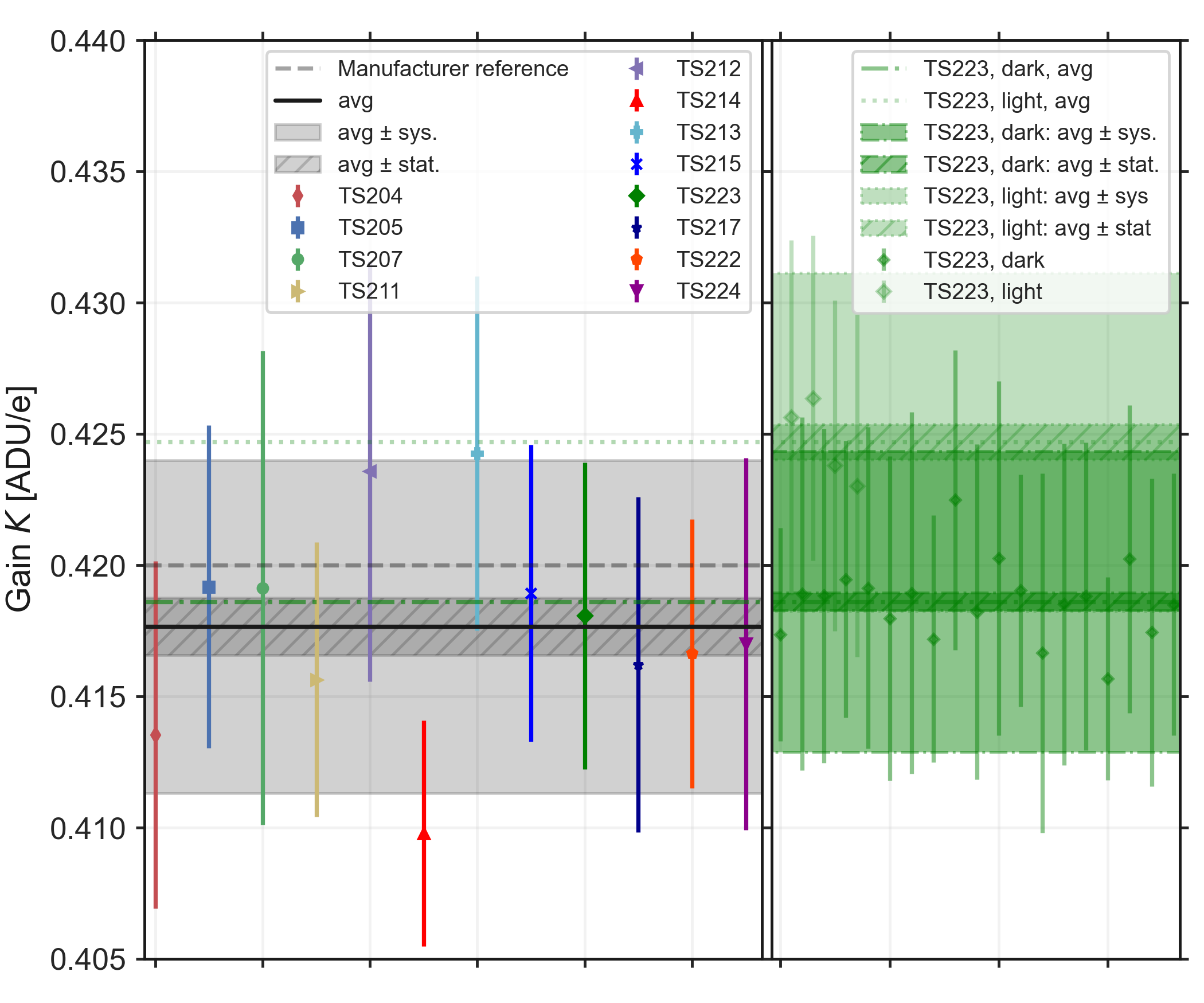} &
                   \includegraphics[height=6.7cm]{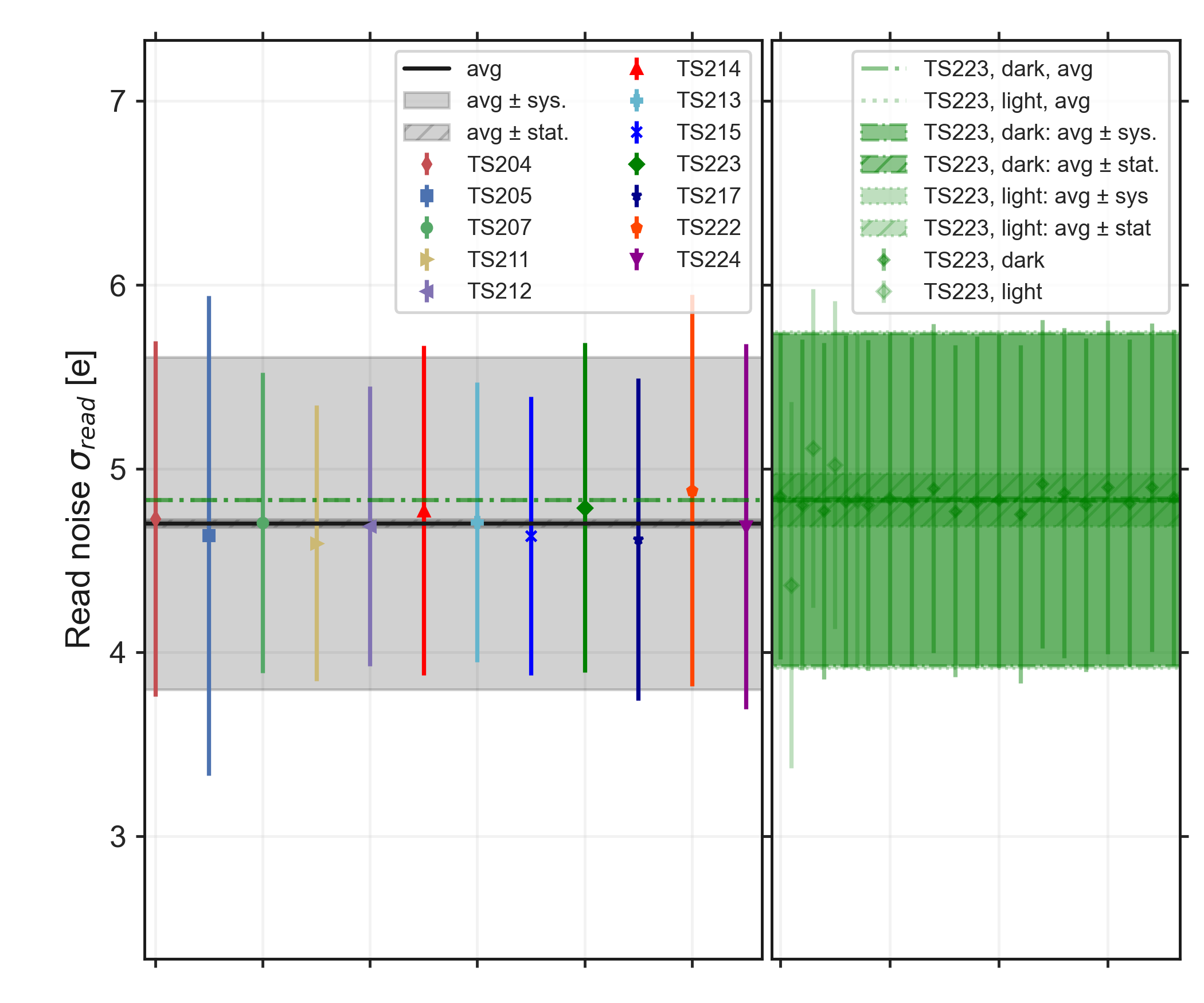}
               \end{tabular}
            \end{center}
            \caption[example] 
            {\label{fig:dc_gain_all_sensors}
            The measurement results for gain (left) and sensor read noise (right) of all test sensors under investigation are shown. On both diagrams, the individual measurement results are compared in the left panel, whereas in the right panel, multiple repeated measurements are shown for the test sensor with ID \emph{TS223} to illustrate the reproducibility. Moreover, in the left panel, the results are compared for dark images and also light images, where the test sensor is illuminated with monochromatic light ($500\,\text{nm} \leq\lambda\leq 900\,\text{nm}$). The average values are indicated by horizontal lines together with bands representing the systematic and statistical uncertainties (Tab.~\ref{tab:dc_results}). The manufacturer reference of the test sensor's gain at $0.42\,\text{ADU}/\text{e}$ is shown by the horizontal dashed line on the left-hand side.}
        \end{figure}
        \begin{table}[ht]
            \caption{Averaged measurement results of the gain measurement procedure} 
            \label{tab:dc_results}
            \small{
            \begin{center}       
                \begin{tabular}{|l|c|} 
                    \hline
                    \rule[-1ex]{0pt}{3.5ex}  Gain $K\,[\text{ADU/e}]$ & $0.418 \pm 0.001_{\mathrm{stat}}\, \pm 0.006_{\mathrm{sys}}$  \\
                    \hline
                    \rule[-1ex]{0pt}{3.5ex}  Read noise $\sigma_{\text{read}}\,[\text{e}]$ & $4.70 \pm 0.02_{\mathrm{stat}}\, \pm 0.90_{\mathrm{sys}}$\\
                    \hline
                \end{tabular}
            \end{center}}
        \end{table}

\begin{figure} [h!]
            \begin{center}
               \begin{tabular}{c}
                   \includegraphics[height=6.5cm]{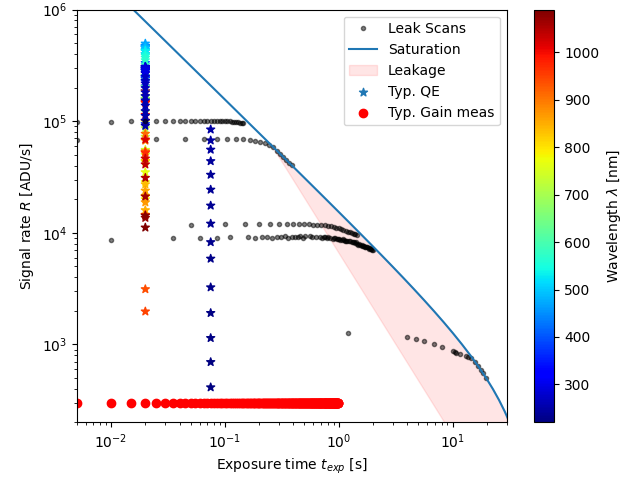}
               \end{tabular}
            \end{center}
            \caption[example] 
            {\label{fig:non_lin} 
            Measured linearity for a single sensor. We estimate the test sensor's signal rate summed over all pixels with varying exposure times with different flux and varying flux but constant exposure time. The signal rate becomes non-linear when some pixel reaches their ADC's saturation limit and if the transfer gate leakage current becomes non-negligible. In the last case, the signal rate follows a power-law. We set the transition to the leakage-dominated area (red shaded) as the intersection of the power-law trend with the former signal rate in the linear regime at constant flux levels. The typical signal rates during the measurements of quantum efficiency (color-coded stars) and gain (red circles) are included in the diagram.}
        \end{figure}

        To determine non-linearity and count rate non-linearity we illuminate the complete pixel array homogeneously with a defocused image of the multimode fiber output. Exposure time sequences are conducted with randomly sorted exposure times for different flux levels using the ND-filter of the optical setup. Any variation in the light source's intensity results in a stronger normal distributed residuals, but does not affect the later linear fit. For each exposure time we take several light and dark images from which a region of 25x25 pixels is used to calculate averaged values. We take the median over all exposures at a fixed exposure time to reject outliers in the signal and background images. A background correction is applied by removing the median background level from the median signal level. We then divide the corrected signal level by the exposure time and obtain an estimate of the rate at which the signal was generated (Fig.~\ref{fig:non_lin}). For an ideal detector the rate estimate should be constant till the pixel approaches saturation or the ADC reaches its voltage range. However, for low rates we find the rate estimates to deviate from a constant earlier than saturation. We attribute this to transfer gate leakage, which is confirmed by the manufacturer as a feature of the test sensors. However, we can also see that our QE measurements (indicated by stars in Fig.~\ref{fig:non_lin}) are far away from these measurement conditions and are therefore not affected by leakage resulting in signal rate non-linearity.
    
    \subsection{Dark current}
    \label{sec:dark_current}
       Tower semiconductor manufactured the test sensors that DESY received,  with the same BSI process which will be used in the final ULTRASAT sensor. Therefore, the test sensor has a similar epitaxial layer to the ULTRASAT sensor. Even though, the pixel dimensions were different, we expect similar dark performance from the test sensor compared with the ULTRASAT flight sensor. The dark signal level of the ULTRASAT sensor at the operational temperature ($-70^{\circ}$C) shall be below 0.026 e/pix/s according to the requirement~\cite{Asif_2021}. It is one of the crucial design parameters of ULTRASAT that is directly related to the mission performance. Therefore, we were particularly interested in the dark signal level at the operating temperature and the dark current doubling factor\cite{EMVA1288} of the test sensor. 

        The experiment setup for the dark current-temperature dependency investigation consists of a thermal enclosure, a temperature monitoring system, the test sensor, and an evaluation board to readout the test sensor (see Fig.~\ref{fig:sens_pack_devboard}). We use the thermal enclosure to cool down the test sensor to the required temperature and monitor the temperature of the test sensor during the experiment. The dark current at the different temperatures is estimated as described in Section~\ref{sec:gain_non_linearity_and_leakage_current}.

             \begin{figure} [h!]
               \begin{center}
                   \begin{tabular}{c}
                       \includegraphics[height=7cm]{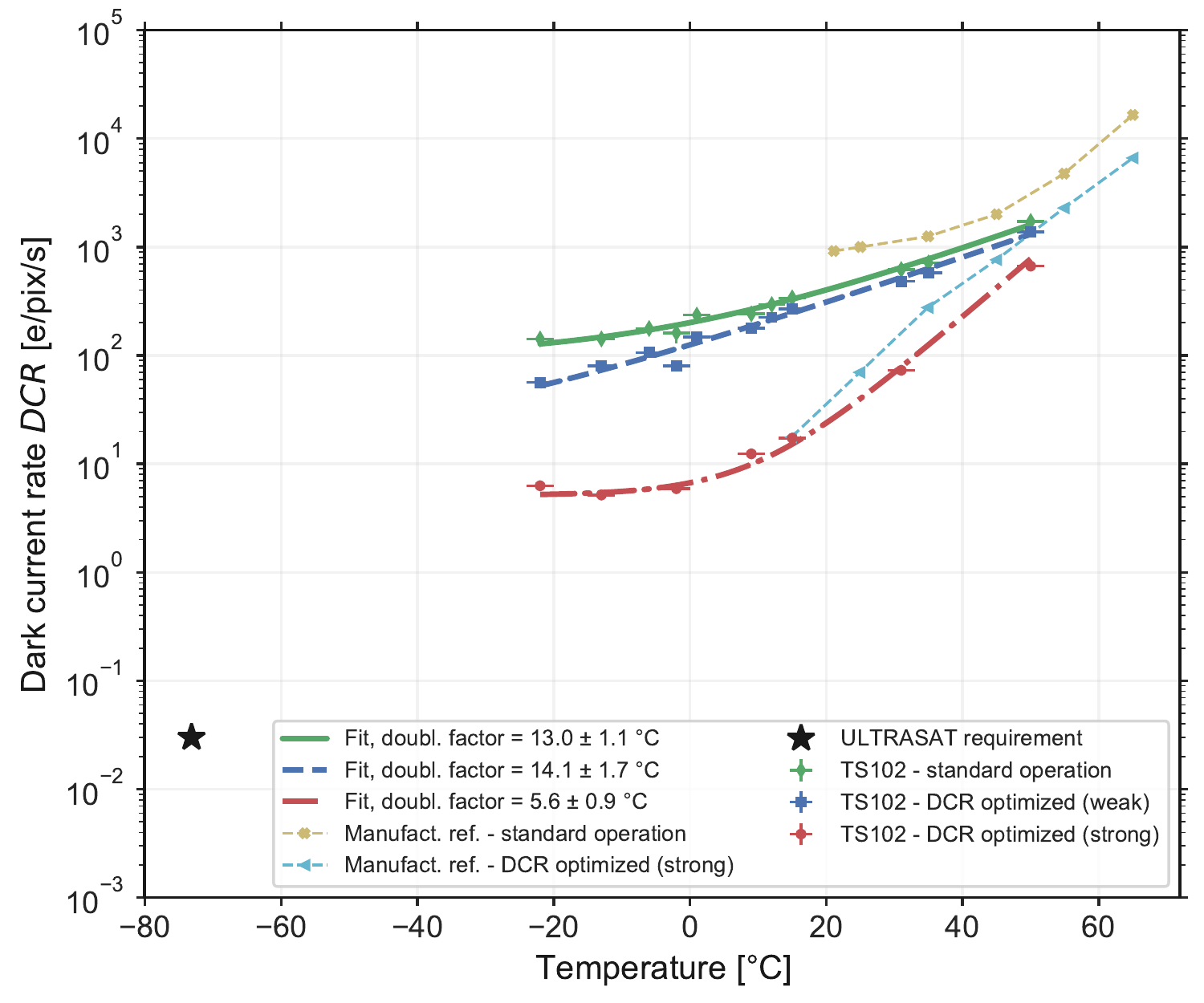}
                   \end{tabular}
                \end{center}
                \caption[example] 
                {\label{fig:dc_results_dc_vs_temp} 
                The measured dark current rate of test sensor \emph{TS102} given in electrons per second an pixel is shown as a function of temperature. From a power-law best-fit estimate the dark current doubling factor is determined. Bias voltages are varied to optimize the DCR performance. The measurement with strong DCR optimization (red, circular markers) shows that the test sensor is capable to meet the ULTRASAT requirement (star marker) if unimpeded by other noise contributions than thermal dark current. However, only the green and yellow measurements represent conditions under which the test sensor is designed to be fully operational.}
            \end{figure} 
        Figure.~\ref{fig:dc_results_dc_vs_temp} shows the results of our dark current measurement campaigns. Above $20^{\circ}$C the dark current decreased steeply with temperature. However, below $20^{\circ}$C dark current didn't follow the previous trend. Further reduction in the temperature didn't reduce the dark current as expected. We suspected the following reasons behind these dark current curves. First, the test sensor gets self-heated during operation. Second, there is a source (external or internal) of dark signal in addition to the usual thermal excitation in the conversion layer.\\
        To infer the impact by self-heating  on the dark current of the test sensor, we used extension cables between the socket and mainboard (Fig~\ref{fig:sens_pack_devboard}) to insulate the test sensor from all the heating elements in the evaluation board. We observed a slight reduction in the dark current, but the issue persisted.

       \begin{figure}[ht]
            \begin{center}
               \begin{tabular}{cc}
                   \includegraphics[height=6.5cm]{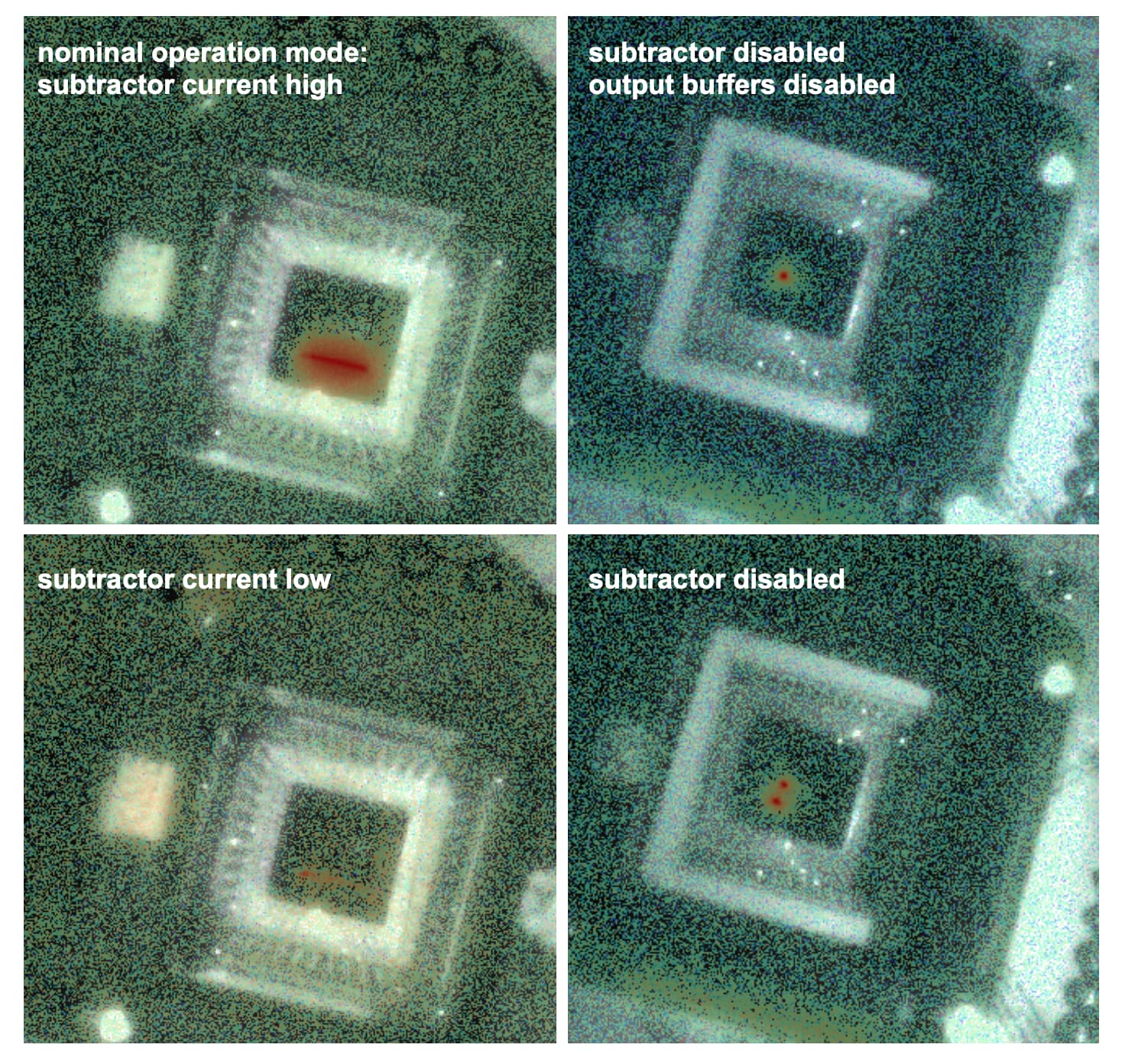} &
                   \includegraphics[height=6.5cm]{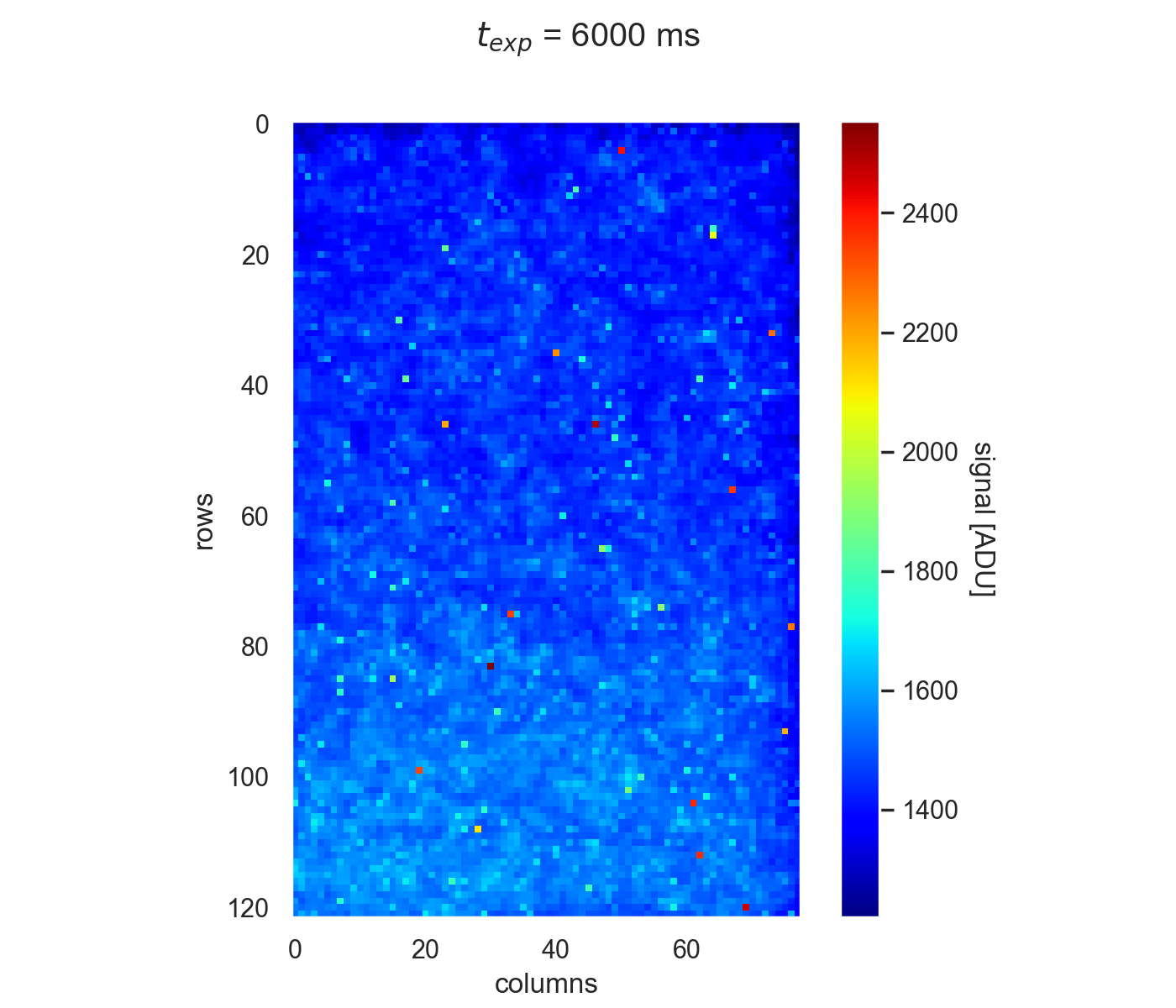} 
               \end{tabular}
            \end{center}
            \caption[example] 
            {\label{fig:dc_ir_emission} 
            Four images (Left) taken with an IR-sensitive camera show a test sensor under different operating conditions with varied parameters of the voltage supplies for the on-die analog signal processing blocks. The upper left image represents the nominal operating condition. The images are superimposed onto a base image taken under illumination from a $950\,$nm LED. A dark image (Right) captured with a test sensor under nominal operating conditions. A rising signal gradient from top to bottom of the image is clearly visible.
            }
        \end{figure} 
        At this point, we studied each dark image separately and observed a gradient in signal level across the image. The lower part of the dark image has comparatively higher signal count than the upper part (refer Fig.~\ref{fig:dc_ir_emission}, right). The non-uniformity of the dark signal suggested a source situated at the bottom of the test sensor pixel array. We rejected the idea of a source external to the test sensor since the freezer is lightproof. We therefore identified IR emission of the periphery electronics on the test sensor's die as the likely source and verified this by applying different bias voltages. For some of the applied bias voltage, we found that the gradient disappeared. This lead to the conclusion that the source is internal.
    
\begin{figure} [h!]
           \begin{center}
               \begin{tabular}{c}
                   \includegraphics[height=6cm]{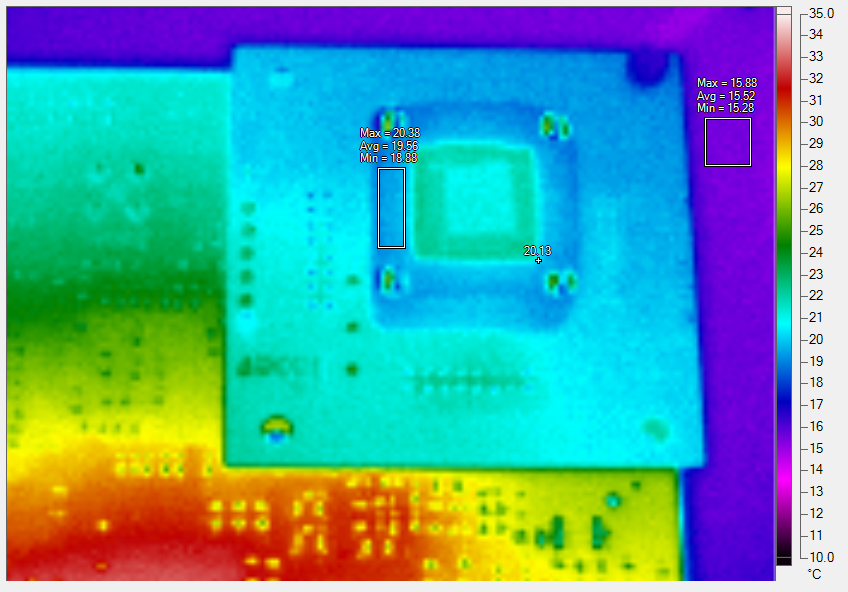}
               \end{tabular}
            \end{center}
            \caption[example] 
            {\label{fig:self_heating} 
            Infrared image of the test sensor and its read out electronic during operation made with a Fluke Ti25 Thermal Imager. The electronics heat up to $\approx 10\,^{\circ}$C higher compared with the test sensor itself (red). The sensor is $\approx 7\,^{\circ}$C warmer than the surrounding environment. The intensity and the exact temperature distribution vary depending on the operation parameters.}
        \end{figure}

        To verify the readout electronics IR emission, we imaged the test sensor using ASI183MMPro camera from ZWO\footnote[8]{https://astronomy-imaging-camera.com/product/asi183mm-pro-mono} while the test sensor is operating. Tower semiconductor has provided us with the register values with which we could turn on or off different parts of the test sensor readout electronics. We imaged the test sensor during its nominal operating mode and while some of the subsystems of the readout electronics turned off or kept at minimal operational condition.  Figure.~\ref{fig:dc_ir_emission} (left) is created by superimposing IR emission image with white light image of the test sensor during different working conditions. Figure.~\ref{fig:self_heating} confirmed our suspicion about the readout circuitry IR emission. The intensity and the location of the IR emission were also different depending upon the mode of operation of the test sensor.  Based on these findings several mitigation methods were implemented into the design of the ULTRASAT sensor, on which we will report in future publications.

    \subsection{Quantum yield}
    \label{sec:quantum_yield}
    
    Quantum yield (QY) is the average number of electron-hole pairs generated in the silicon photodetector per photon interacting with the silicon. Studies \cite{Blake2011, Canfield_1998} have shown that high-energy photons with a wavelength less than $300\,$nm will generate more than one electron-hole pair in the silicon. At the lower wavelength regime, QY might lead to the overestimation of the Quantum Efficiency (QE). QE is the photon to electron conversion probability in silicon photodetectors conventionally measured as the ratio of the collected electrons to the incident photons. Thus, one might overestimate the QE below $300\,$nm because of the additionally created electrons. Therefore, the QY is the correction factor to the overestimated QE. We measure the QY in the operational bandwidth of ULTRASAT ($220\,$nm - $280\,$nm) for verification and completeness of our measurement even though previous studies and models were available.

    We follow the method by Janesick et al.~\cite{Janesick1987} to estimate the QY. We define the QY as
    \begin{equation}
            \mathrm{QY(\lambda)}  =  \frac{J(\lambda)}{K(\lambda > 400\,\mathrm{nm})} \, ,
            \label{eq:qy}
        \end{equation}
    where $J$ is the gain estimated from the light images with a wavelength less than $400\,$nm and $K$ is the gain estimated from the low energy photon light images with a wavelength above $400\,$nm for normalization.

    We generate a full photon transfer curve (PTC) at every $5\,$nm from $220\,$nm to $280\,$nm by illuminating the test sensor completely with the monochromatic light. The data analysis follows the procedure described in Section~\ref{sec:gain_non_linearity_and_leakage_current}. The full width at half maximum of the monochromatic light used is $3.4\,\text{nm}$. For normalization, we estimate the gain $K$ at $500\,$nm, $700\,$nm and $900\,$nm.

\begin{figure} [h!]
           \begin{center}
               \begin{tabular}{c}
                   \includegraphics[height=6.5cm]{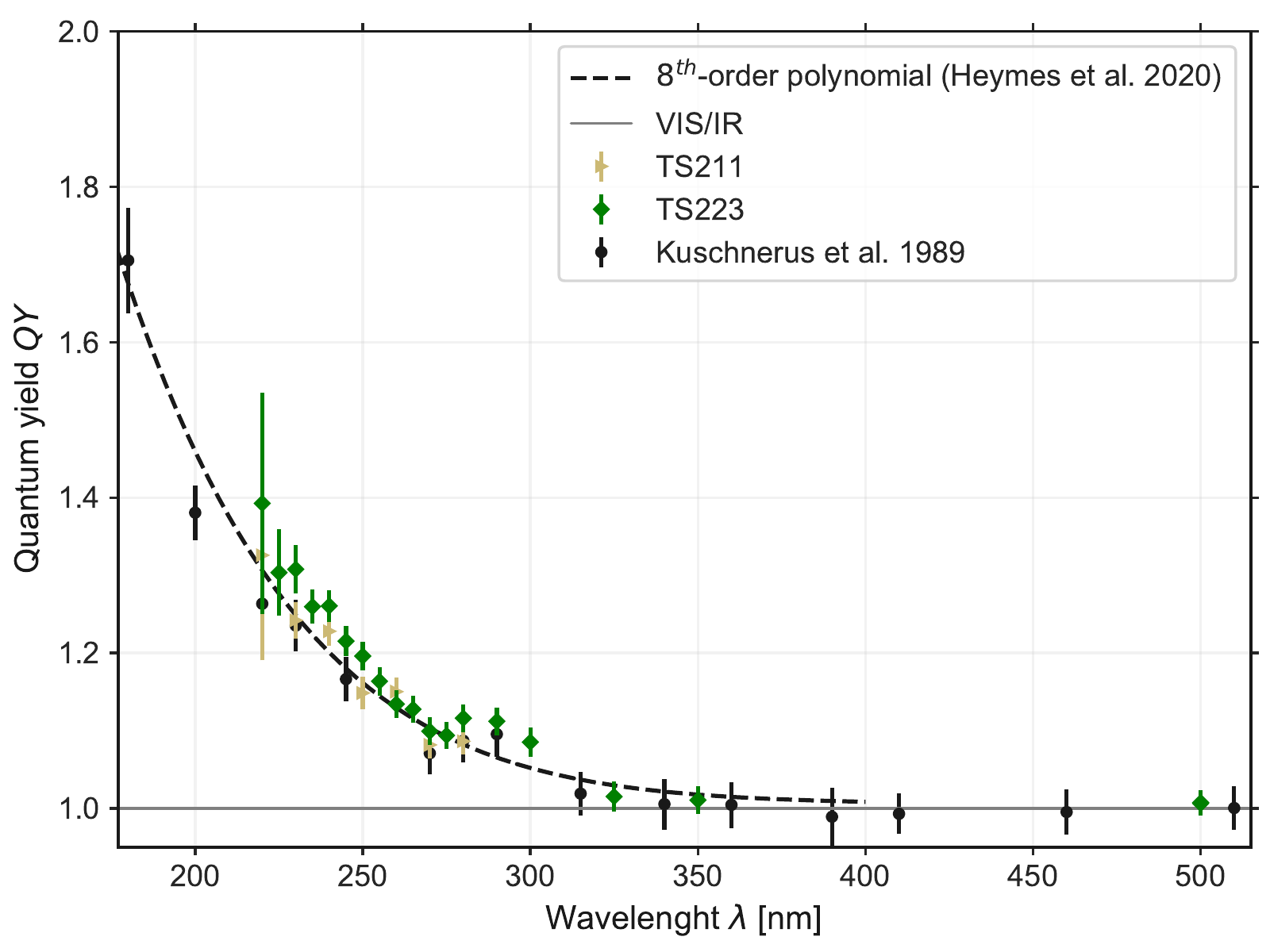}
               \end{tabular}
            \end{center}
            \caption[example] 
            {\label{fig:qy_results} 
            The green and beige markers show test sensor measurements of quantum yield as a function of wavelength. The measured values are normalized to the average measured light gain between $500\,\mathrm{nm}$ and $900\,\mathrm{nm}$ (Fig.~\ref{fig:dc_gain_all_sensors}). The solid line at $QY = 1$ represents the theoretical quantum yield expected for light in the visual (VIS) and infrared (IR) spectrum. For comparison, the black circles show the reproduced quantum yield data based on direct measurements in silicon (see description in text)~\cite{kuschnerus1998characterization}. From this data a best-fit model for quantum yield (eighth-order polynomial)~\cite{Heymes2020} is derived, represented by the dashed line. The model is found to be valid for light below $400\,\text{nm}$ and is used in this study to compute the wavelength-dependent quantum yield correction factor.}
        \end{figure}

    Figure~\ref{fig:qy_results} shows the measured QY for two test sensors in the wavelength range from $220\,$nm to $500\,$nm. We follow Heymes et al. to empirically model the QY~\cite{Heymes2020}. The authors use a theoretical description of the QY based on measured data on the ionisation energy of silicon\cite{kuschnerus1998characterization} to derive an eighth-order polynomial fit between $40\,$nm to $400\,$nm. We use this description to reproduce the QY data on which the fit is based. The comparison to our measured QY shows the two data sets are in good agreement. Thus, we take this polynomial to model the QY and to correct the QE measurements for wavelengths below $400\,\mathrm{nm}$. An overall systematic uncertainty of $2\,\%$ on the QY derived from this model is assumed.\cite{kuschnerus1998characterization}

    \subsection{Quantum efficiency}
    \label{sec:quantum_efficiency}
        
        The measurement of the spectral quantum efficiency\footnote[9]{We use the definition of the interacting quantum efficiency as it is defined in } is a crucial part in this characterization as it influences the decision on the anti-reflective coating (ARC) for the final UV sensor of ULTRASAT.
        
        We measure the quantum efficiency $QE(\lambda)$ as function of the wavelength $\lambda$. To illuminate the test sensor we use monochromatic light with a bandwidth of $3.4\,$nm below $365\,$nm and $1.7\,$nm above\footnote[1]{Diffracting gratings with $600\,\mathrm{grv}\,\mathrm{mm}^{-1}$ lead to an increase in the flux compared to $1200\,\mathrm{g}\,\mathrm{mm}^{-1}$ (Sec.~\ref{sec:performance_results})}. The QE can be estimated by comparison of the flux of incident photons per s $F_{\gamma}(\lambda)$ with the test sensor's signal response $S$ in photoelectrons per s. The QE can be expressed as:
        \begin{equation}
        	\mathrm{QE}(\lambda) = \frac{S}{\mathrm{QY}(\lambda)\,F_{\gamma}}\,,
        	\label{eq:spectral_qe}
        \end{equation}
        with the wavelength-dependent quantum yield $\mathrm{QY}(\lambda)$. To convert ADUs into photoelectrons we use the gain $K$ from Table~\ref{tab:dc_results} and the exposure time $t_{\mathrm{exp}}$: $S[\si{\electron\per\second}] = S[\si{\adu\per\second}] (K[\si{\adu\per\electron}]\,t_{\mathrm{exp}}[\si{\second}])^{-1}$.
        
        For each data point we collect a number of images on the order of $O(10)$ and measure the photocurrent of the working standard simultaneously. We perform two background measurements each time, one before illumination and one after, and subtract the averaged background from signal. The test sensor is illuminated by a circular shaped beam spot with a FWHM $(530\pm30)\,\mathrm{\mu m}$ which covers $\approx 1000$\,pixels, or $\approx 10\,\%$ of the test sensor's collective area. The setup allows to vary the angle of incidence (AoI). With the variation in the wavelength the focus is being corrected automatically by the gantry robot following the lens equation and the Sellmeier dispersion equation. 

\begin{figure} [ht]
            \begin{center}
               \begin{tabular}{cc}
                   \includegraphics[height=6cm]{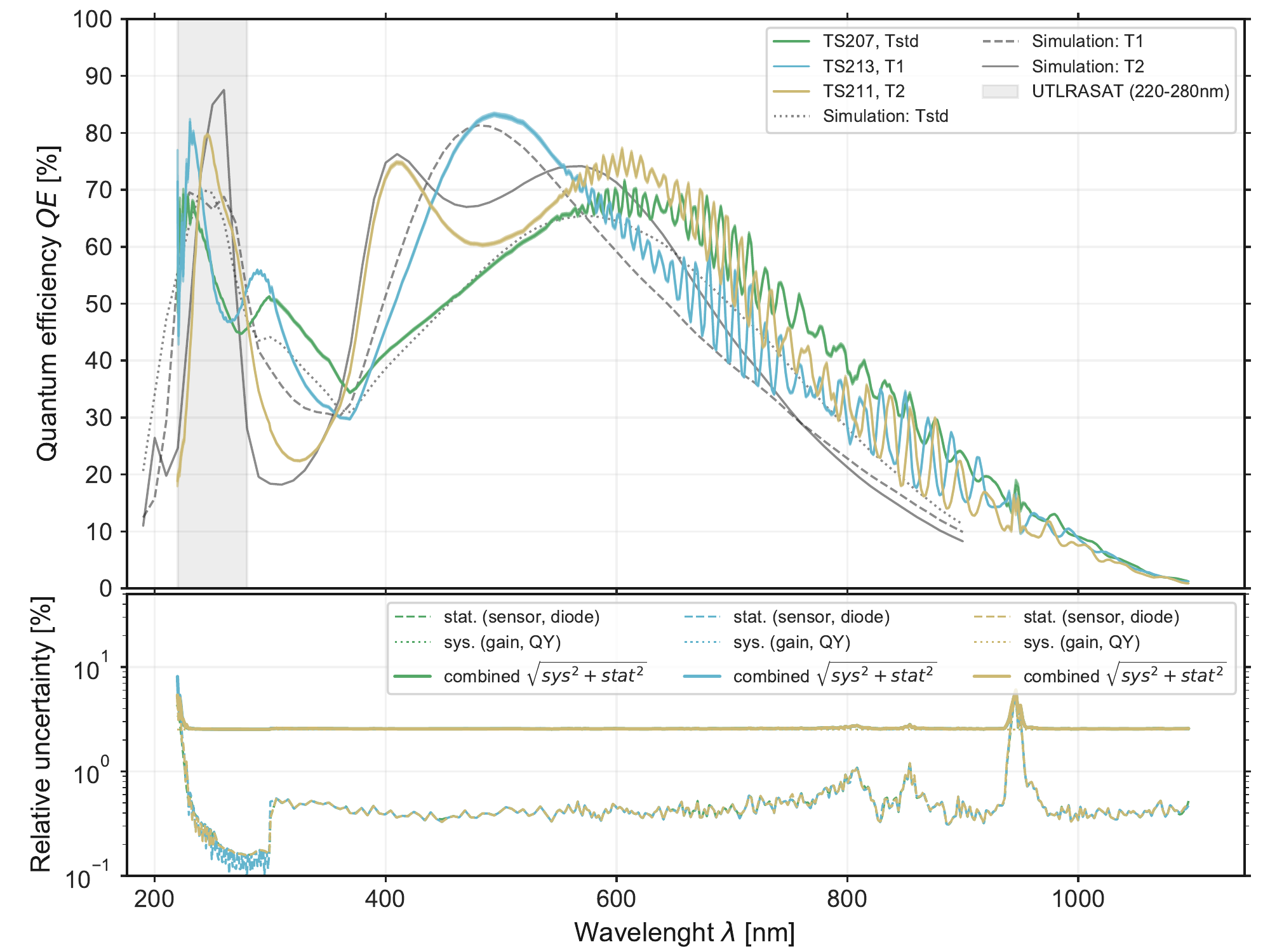} & \includegraphics[height=6cm]{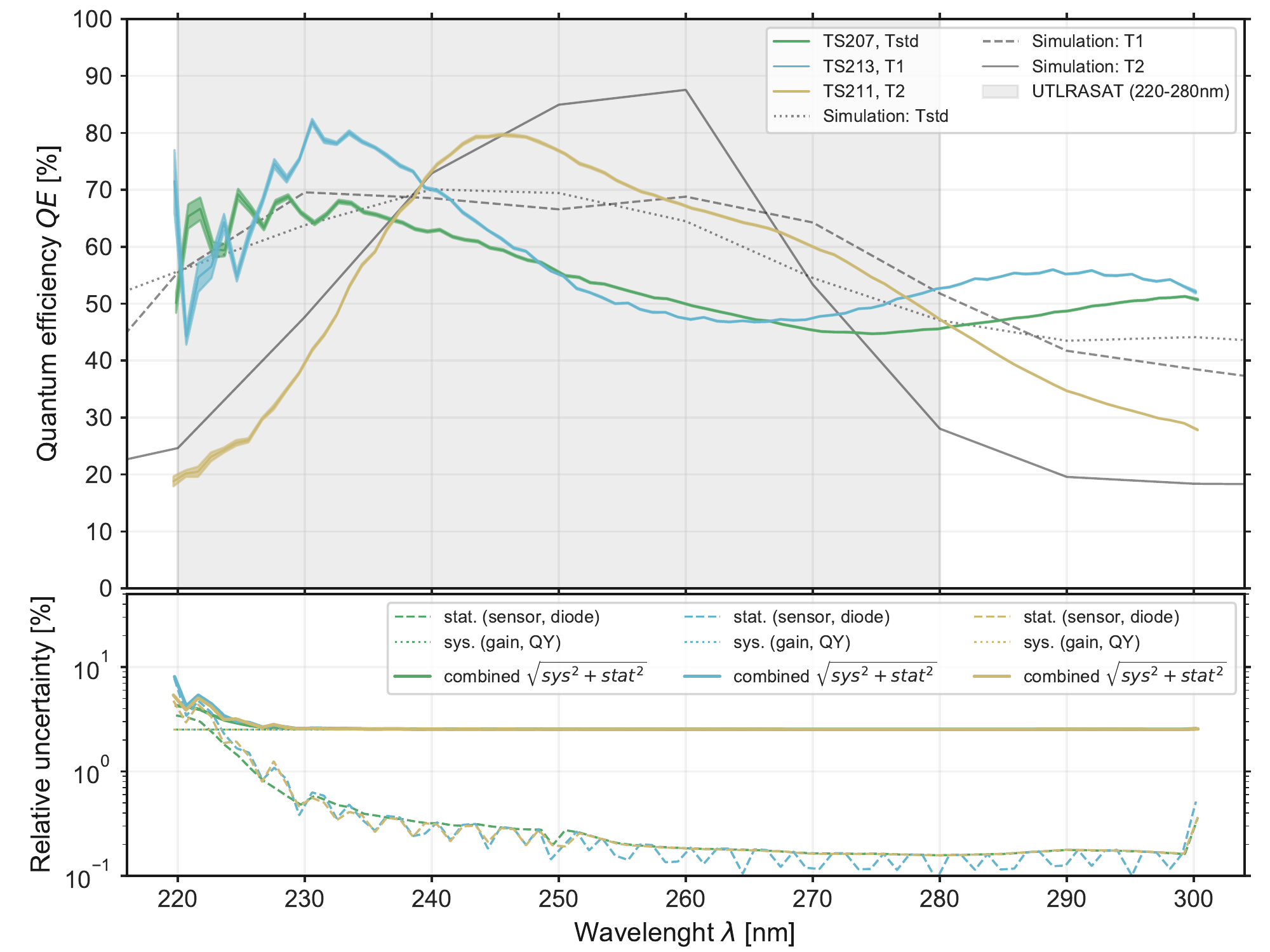} 
               \end{tabular}
            \end{center}
            \caption[example] 
            {\label{fig:qe_reference} 
            The spectral quantum efficiency for the three ARC options is compared across the entire spectral range (left) and the operational waveband (right) of ULTRASAT.
            }
        \end{figure}

        We measured the QE of test sensors of all three ARC options (Fig~\ref{fig:qe_sim}) with a resolution of $1\,$nm in\\ $[220\,\mathrm{nm}, 300\,\mathrm{nm}]$ and $2\,$nm in $[300\,\mathrm{nm}, 1100\,\mathrm{nm}]$. In both UV and IR spectral bands we observe fringing. Some characteristic values within the ULTRASAT operational waveband of these measurements are summarized in Tab.~\ref{tab:qe_results}. To test reproducibility, we perform additional measurements on different test sensors of the same ARC option as well as repeated scans of single sensors in the time frame between October 2020 and July 2021. The resolution for if  the reproducibility measurements is typically larger with $5\,$nm and $10\,$nm in the UV and visual/IR bands, respectively. Figure~\ref{fig:qe_reference} shows QE measuruement results of representative test sensors of all three ARC options together with their corresponding simulations. The measurements are preformed with an AoI of $20\,^{\circ}$, whereas fro the simulated data an AoI of $\approx 25\,^{\circ}$ is assumed. The left-hand side plot shows the measurement across the full spectral range of the setup from $200\,$nm to $1100\,$nm. The measured QE follows qualitatively the simulations' trends. As the measurements are performed with a smaller AoI it is shifted red against the simulations. While coating option Tstd and T1 do not clearly decline towards the far-UV-end of the ULTRASAT operational waveband, T2 rejects the far-UV more effectively and peaks in the center at $\approx 245\text{nm}$ of the waveband. At a wavelength at $220\,\mathrm{nm}$ the efficiency is  $(19\pm1)\,\%$ and its maximum is $(80\pm2)\,\%$.
        
        \begin{table}[ht]
            \caption{Characteristic values from the measured QE of the three  high-resolution scans in the ULTRASAT operational waveband (Fig.~\ref{fig:qe_reference})} 
            \label{tab:qe_results}
            \small{
            \begin{center}       
                \begin{tabular}{|c|c|c|c|c|c|} 
                    \hline
                    \rule[-1ex]{0pt}{3.5ex} ARC & Test Sensor ID & $\text{QE}(220\,\text{nm})\,[\%]$ & \text{QE}(250\,\text{nm})\,[\%] & \text{QE}(280\,\text{nm})\,[\%] & Averaged QE $[\%]$ \\
                    \hline
                    \rule[-1ex]{0pt}{3.5ex}  Tstd & TS207 & $50 \pm 2 $& $56 \pm 2$ & $46 \pm 2$ & $56 \pm 1$\\
                    \hline
                    \rule[-1ex]{0pt}{3.5ex}  T1  & TS213 & $72 \pm 6 $ & $56 \pm 1$ & $53 \pm 1$ & $58 \pm 2$\\
                    \hline
                    \rule[-1ex]{0pt}{3.5ex}  T2  & TS211 & $19 \pm 1 $ & $77 \pm 2$ & $48 \pm 1$ & $58 \pm 2$\\
                    \hline
                \end{tabular}
            \end{center}}
        \end{table}

\begin{figure} [H]
            \begin{center}
               \begin{tabular}{c}
                 \includegraphics[height=6.5cm]{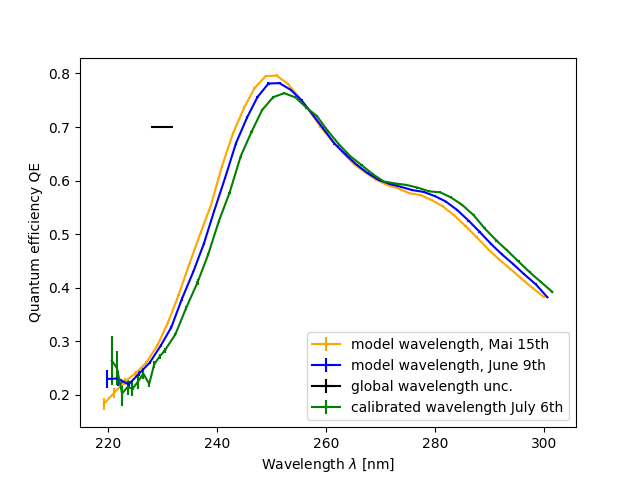}
               \end{tabular}
            \end{center}
            \caption[example] 
            {\label{fig:qe_reproducibility} 
            Shown are three QE scans of the same test sensor. The measurements are performed in the wavelength range 220-300\,nm and over a period of 3 month. The first two measurements are subject to large wavelength uncertainties shown by a black errorbar in the upper left. The last measurement is calibrated using a Holmium Didymium absorption line filter leading to 0.2\,nm uncertainties on the measurement.
            }
        \end{figure}

        To study systematic changes due to the storage of the test sensor under ambient air without dedicated cleanliness monitoring, we performed several measurements on one sensor over three months. Figure~\ref{fig:qe_reproducibility} shows three repeated scans of the same test sensor within the wavelength range 220-300\,nm.
        The uncertainty on the wavelength is different for the three scans. For the first and second scan in May and June the Holmium Didymium absorption line filter was not available, therefore, the uncertainty is $\approx$\,2\,nm. For the last scan we used the absorption lines to calibrate the measurement to the filter's calibration uncertainty of 0.2\,nm provided by Hellma (compare Section~\ref{sec:performance_results}). 
        We observe a systematic shift of the QE peak wavelength with time, but due to the lack of a filter during the first and second measurements and the resulting larger uncertainties, we do not consider this shift to be significant.
        We also see a change in peak amplitude. There are two potential reasons for this, a true change of the test sensor under ambient air or an artifact due to systematic differences in the wavelength of the QE measurements and the QY measurements.
        We will continue with these measurements to infer potential sensor degridation and starting from July all further measurements will be referenced to the absorption lines in the Holmium Didymium filter. The observed maximal variation on the peak QE within the ULTRASAT operational waveband is $\approx 5\,\%_{\mathrm{abs}}$ for this individual sensor over the time. The deviation between different test sensors of the same coating option is in the same order of magnitude.
        
        \begin{figure} [h!]
            \begin{center}
               \begin{tabular}{c}
                 \includegraphics[height=6cm]{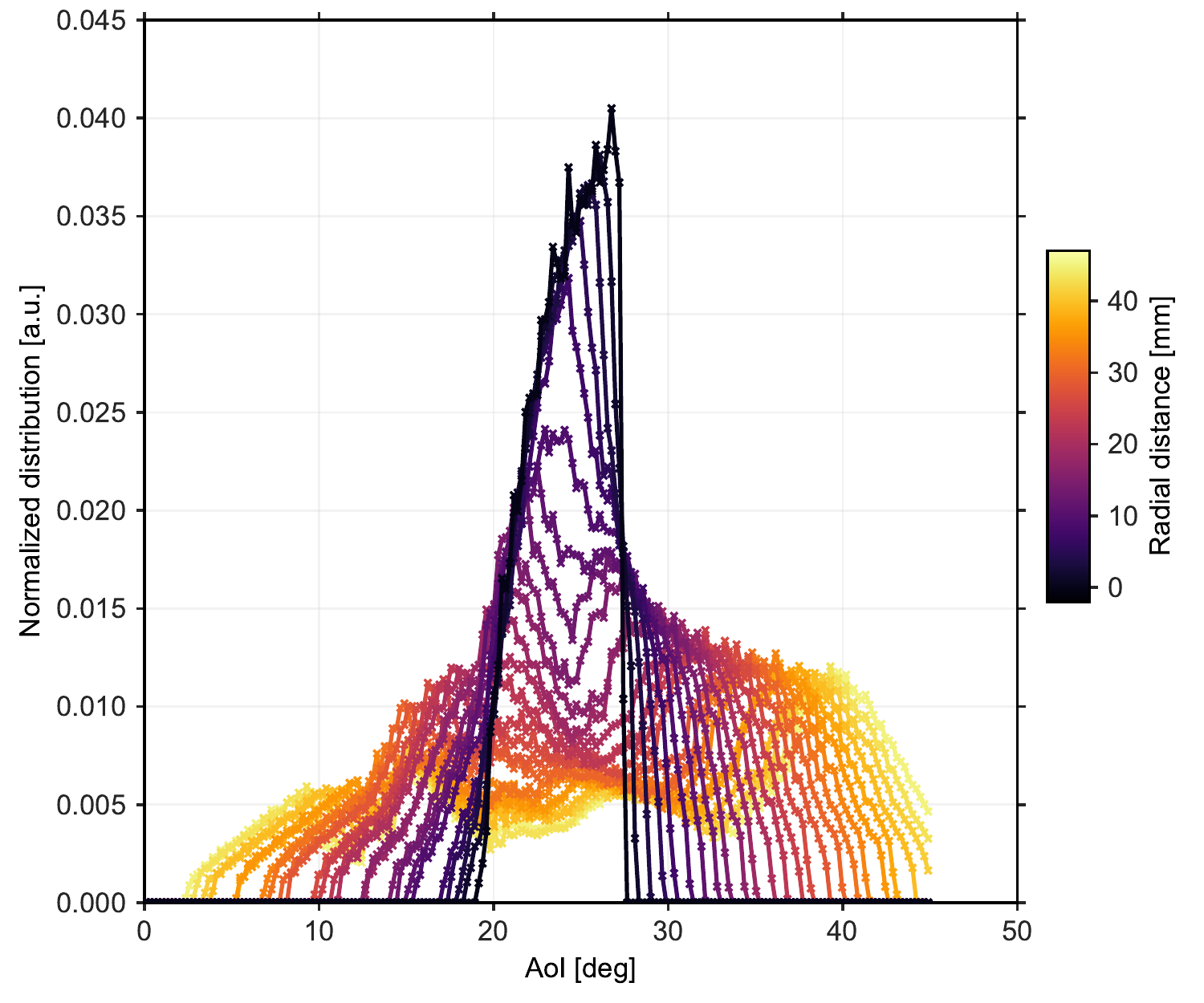}
               \end{tabular}
            \end{center}
            \caption[example] 
            {\label{fig:qe_aoi_dist}
            The normalized distributions of angle of incidence at the focal plane of the ULTRASAT telescope are shown color-coded with respect to the corresponding radial distance to the optical axis. The shown data is obtained from numerical simulations of ULTRASAT's telescope optics.
            }
        \end{figure}
        \begin{figure} [h!]
            \begin{center}
               \begin{tabular}{c}
               \includegraphics[height=5cm]{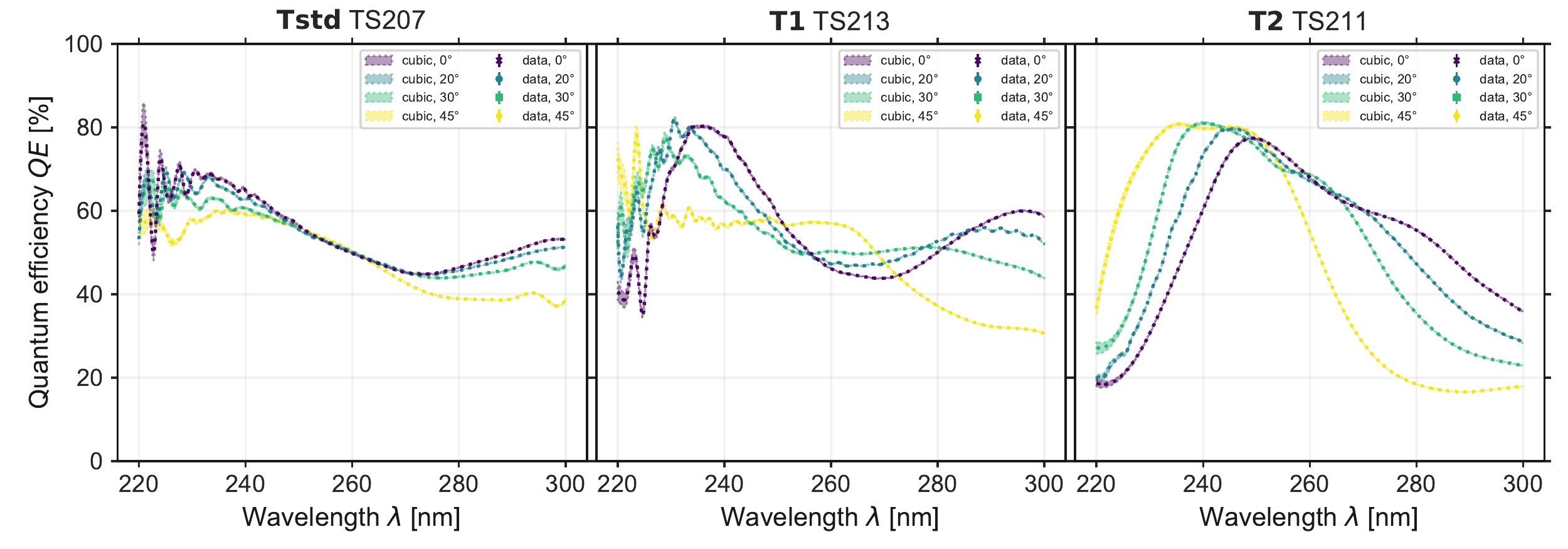}\\ \includegraphics[height=5cm]{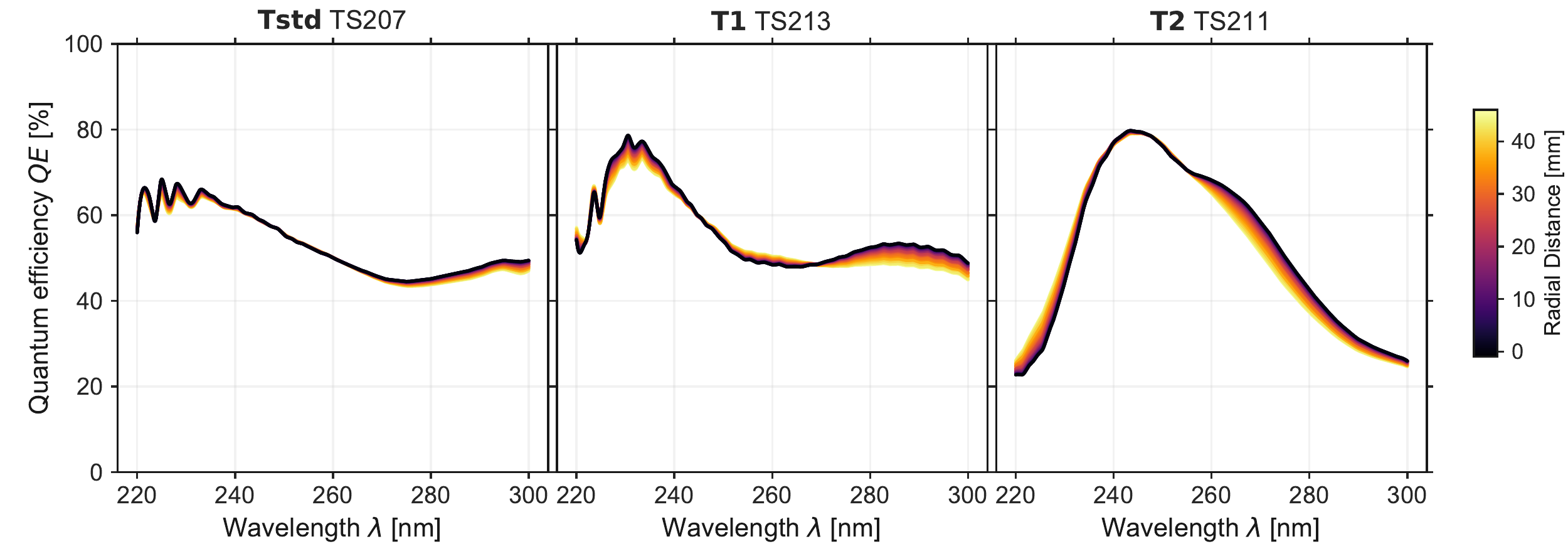} 
               \end{tabular}
            \end{center}
            \caption[example] 
            {\label{fig:qe_angular_uv} 
            AoI comparison of $QE(\lambda)$ in the ULTRASAT operational waveband extended to $300\,\text{nm}$. Shown is the measured data at four different AoIs with cubic spline interpolation (left). The $QE(\lambda)$, weighted by the distribution of angle of incidence (Fig.~\ref{fig:qe_aoi_dist}), are shown color-coded with respect to the radial distance from the telescope's optical axis (right).
            }
        \end{figure}

        Due to the obscuration and the telescope optics, the angle of incidence (AoI) has a non-trivial distribution at the final sensor which is dependent on the radial distance from the center of the focal plane. The transmission of ARC and therefore the sensor's response differs with the AoI. To quantify this effect on the QE we measure the spectral quantum efficiency with varying inclination angles in the region of interest $[0\,^{\circ}, 45\,^{\circ}$] relative to the sensor's normal. The test sensor was mounted on a rotational platform to rotate it with respect to incident light beam (Fig.~\ref{fig:qe_setup}). It is clearly shown in Figure.~\ref{fig:qe_angular_uv} that the measured QE is shifted blue. For inclined light rays the phase difference between reflected and transmitted rays in the ARC is reduced with increasing AoI.\cite{quijada2004angle} Furthermore, we weight the spectral QE with the expected AoI distribution as a function of the radial distance from the telescope's optical axis (Fig.~\ref{fig:qe_aoi_dist}). The result is shown on the right-hand side of Figure.~\ref{fig:qe_angular_uv}. The AoI-weighted QE form a set of curves whereby the measurement with an incident angle $25\,^{\circ}$ can be identified as its representative average. The set varies maximal $\pm 5\,\%_{\mathrm{abs}}$ from this representative. The reason being that the decreasing spectral response on one side of the ULTRASAT operational waveband is compensated by an increasing response on the other side and vice versa. Hence, we conclude the AoI-dependence has a negligible effect on the general QE performance and especially when compared to the impact by the ARC.

        We estimated the spectral quantum efficiency of the test sensors by comparing its signal response with the calibrated light flux estimated by our working standard. The relative uncertainties for these efficiencies are limited by the systematic uncertainties on the gain and quantum yield which is $\approx 2.5\,\%_{\mathrm{rel}}$ in the range from $230\,$nm to $1100\,$nm. Only below $230\,$nm and in a absorption region of the multimode fiber at $\approx 950\,$nm, the relative uncertainty is typically between  $4\,\%_{\mathrm{rel}}$ and $8\,\%_{\mathrm{rel}}$ limited by the statistical uncertainty due to the sparsely available flux of the light source. We tested all three ARC options, estimated the production reproducibility by testing multiple sensors of the same coating and varied the angle of inclination to quantify the effect of the telescope's obscuration and optics on the spectral QE. We observe variations in the peak QE for repeated measurements with single sensors as well as for sensors with the same ARC of up to $\approx 5\,\%_{\mathrm{abs}}$. The telescope's overall performance is sensitive to the degree of out-of-band rejection. Out of the three ARCS options under investigation,  option T2 provides the best rejection potential, as it shows a prominent peak-like shape in the middle of the operational band. Its QE shown in Fig.~\ref{fig:qe_angular_uv} peaks at $(79\pm1)\,\%_{\mathrm{abs}}$ and declines for decreasing wavelengths to $(20\pm4)\,\%_{\mathrm{abs}}$ at $220\,$nm. Hence, we recommend the anti-reflective coatings option T2 for the production of the final ULTRASAT UV sensor.

\section{CONCLUSION}
\label{sec:conclusion}
    We presented the results of the test sensor characterization for the ULTRASAT camera. We have established  the methodical procedure which will be applied in future studies of the final ULTRASAT UV sensor. The two main results are:
    \begin{enumerate}
        \item The test sensors show a temperature independent minimal dark current floor at $\approx 7\,$e/s/pix (see Fig.~\ref{fig:dc_results_dc_vs_temp}) We conclude that this is caused by self-heating of the read out electronics and on-die infrared emission (Fig.~\ref{fig:dc_ir_emission}). Although it is not possible to make a definitive statement about the dark current performance of the test sensors at the operating temperature of ULTRASAT's camera, this result led to important design improvements in the final sensors that will be described and verified in upcoming publications. 
        \item From the estimated spectral quantum efficiency of the different ARC options, we expect T2 to optimize the out-of-band rejection of the telescope due to its peak-like shape in ULTRASAT operational waveband (max. QE $\approx 80\,\%$ at $245\,\mathrm{nm}$, Fig.~\ref{fig:qe_reference}). Hence, we recommend ARC option T2 to be chosen for the production of the final ULTRASAT UV sensor.
    \end{enumerate}
    The final sensor is expected to be available from production in early fall 2021. The characterization team is working on increasing the precision of the wavelength calibration as well as software integration of the final sensor into the calibration setup.
    
\acknowledgments
We would like to acknowledge the help and support of the ULTRASAT camera advisory board, composed of Andrei Cacovean, Maria F\"urmetz, Norbert  Kappelmann, Olivier Limousin, Harald Michaelis, Achim Peters, Chris Tenzer, Simone del Togno, Nick Waltham and J\"orn Wilms. We also thank Thorlabs for loaning of the OSA201C Fourier transform spectrometer.  

\bibliography{report} 
\bibliographystyle{spiebib} 

\end{document}